\pdfoutput=1

\documentclass[pdflscape,lscape,iop,apj,numberedappendix,appendixfloats,twocolumn,twocolappendix,stfloats,url]{emulateapj}
\usepackage{paralist,amsmath}
\usepackage{microtype}

\usepackage{txfonts}
\makeatletter
  \newcommand\tinyv{\@setfontsize\tinyv{7.0pt}{7.0}}
\makeatother

\slugcomment{Not to appear in Nonlearned J., 45.}

\shorttitle{Observed Faraday Effects in Damped Lyman-Alpha Absorbers and Lyman Limit Systems}
\shortauthors{Farnes et al.}

\begin{document}


\title{Observed Faraday Effects in Damped Lyman-Alpha Absorbers and Lyman Limit Systems: \\ The Magnetised Environment of Galactic Building Blocks at Redshift=2}


\author{J.~S. Farnes\altaffilmark{1}, L. Rudnick\altaffilmark{2}, B.~M. Gaensler\altaffilmark{3}, M. Haverkorn\altaffilmark{1}, S.~P. O'Sullivan\altaffilmark{4}, S.~J. Curran\altaffilmark{5}}
\altaffiltext{1}{Department of Astrophysics/IMAPP, Radboud University, PO Box 9010, NL-6500 GL Nijmegen, the Netherlands.}
\altaffiltext{2}{Minnesota Institute for Astrophysics, School of Physics and Astronomy, University of Minnesota, 116 Church Street SE, Minneapolis,
MN 55455, USA.}
\altaffiltext{3}{Dunlap Institute for Astronomy and Astrophysics, University of Toronto, ON, M5S 3H4, Canada.}
\altaffiltext{4}{Instituto de Astronomía, Universidad Nacional Autónoma de México (UNAM), A.P. 70-264, 04510 México, D.F., Mexico.}
\altaffiltext{5}{School of Chemical and Physical Sciences, Victoria University of Wellington, PO Box 600, Wellington 6140, New Zealand.}
\email{j.farnes@astro.ru.nl}


\begin{abstract}
Protogalactic environments are typically identified using quasar absorption lines, and these galactic building blocks can manifest as Damped Lyman-Alpha Absorbers (DLAs) and Lyman Limit Systems (LLSs). We use radio observations of Faraday effects to test whether DLAs and LLSs host a magnetised medium, by combining DLA and LLS detections throughout the literature with 1.4~GHz polarization data from the NRAO VLA Sky Survey (NVSS). We obtain a control, a DLA, and a LLS sample consisting of 114, 19, and 27 lines-of-sight respectively -- all of which are polarized at $\ge8\sigma$ to ensure Rician bias is negligible. Using a Bayesian framework, we are unable to detect either coherent or random magnetic fields in DLAs: the regular coherent magnetic fields within the DLAs must be $\le2.8$~$\muup$G, and the lack of depolarization is consistent with the weakly magnetised gas in DLAs being non-turbulent and quiescent. However, we find mild suggestive evidence that LLSs have coherent magnetic fields: after controlling for the redshift-distribution of our data, we find a 71.5\% probability that LLSs have a higher RM than a control sample. We also find strong evidence that LLSs host random magnetic fields, with a 95.5\% probability that LLS lines-of-sight have lower polarized fractions than a control sample. The regular coherent magnetic fields within the LLSs must be $\le2.4$~$\muup$G, and the magnetised gas must be highly turbulent with a typical scale on the order of $\approx5$--20~pc, which is similar to that of the Milky Way. This is consistent with the standard dynamo pedagogy, whereby magnetic fields in protogalaxies increase in coherence and strength as a function of cosmic time. Our results are consistent with a hierarchical galaxy formation scenario, with the DLAs, LLSs, and strong magnesium~II (Mg\,{\sc ii}) systems exploring three different stages of magnetic field evolution in galaxies.
\end{abstract}




\section{Introduction}
\label{introduction}
The evolution of magnetism in galaxies is of fundamental interest \citep{1982ApJ...263..518K}. In particular, the magnetic fields in young protogalaxies remain essentially completely unexplored \citep{1992ApJ...388...17W,1995ApJ...445..624O}. These protogalaxies are expected to have not yet experienced significant dynamo activity, and therefore constitute ``missing links'' in our understanding of the dynamo process and the evolution of cosmic magnetic fields \citep{2004NewAR..48.1003G}. The dynamo process describes how weak ``seed'' magnets in the early Universe were amplified and ordered throughout cosmological history by large-scale rotation and turbulence within galaxy disks and halos \citep[e.g.][]{2014MNRAS.443.1867C,2015ApJ...808...28C}. This suggests that the magnetic field strength should be weaker in protogalaxies than in normal galaxies. Furthermore, the coherence of the typical protogalactic magnetic field should also be less, with significant field disorder expected to be present. Measuring these magnetic fields would have implications for the cosmological growth of magnetism, and would constrain dynamo mechanisms \citep[e.g.][]{2013MNRAS.435.3575B}. Such studies are highly challenging, as at radio wavelengths, the protogalactic environment is typically of such low-luminosity that directly imaging the emission due to magnetic fields will likely not even be possible with ultra-sensitive (sub-$\muup$Jy) data from the Square Kilometre Array (SKA) \citep{2015aska.confE..92J}. We therefore suggest an alternative approach to study magnetic fields in these galactic building blocks now, using radio polarization observations. 
\begin{figure*}
\centering
\includegraphics[trim=0cm 0.0cm 0cm 0.0cm, clip=true, width=15.8cm]{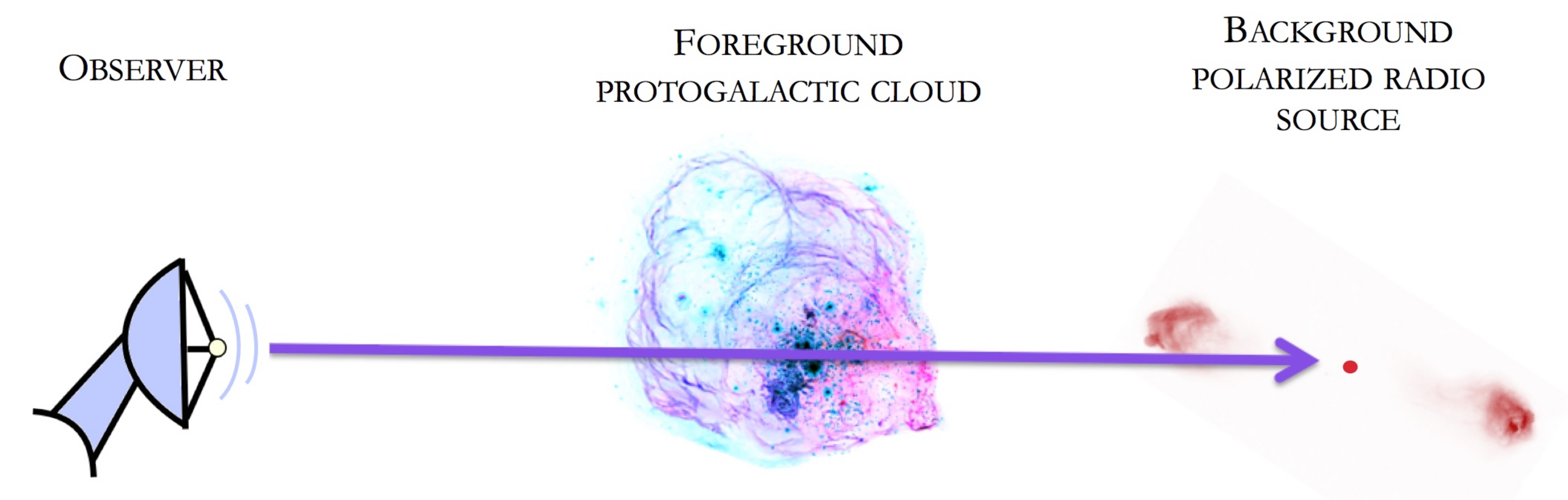}
\caption{A cartoon of a ``back-lit'' experiment towards a polarized background source. This is merely an illustratory cartoon, and is not to scale. The observer, on the left of the cartoon, measures properties (such as Faraday rotation) along the line-of-sight towards a bright, polarized, background radio source. Along the line-of-sight, an intervening foreground cloud is known to be present via quasar absorption lines. The physical nature of the foreground cloud can vary based upon the absorption line used, and in this cartoon is a primeval galaxy. The image is derived from original works in the public domain: the image of the protogalaxy is courtesy of Adolf Schaller for STScI, and the image of Cygnus A is courtesy of NRAO/AUI/NSF.}
\label{DLAcartoon}
\end{figure*}

Radio polarization observations are the best available probe of cosmic magnetic fields, as they allow us to measure both the polarized fraction towards distant background radio sources (which is related to the degree of ordering of magnetic fields), and the Faraday rotation located along the entire line-of-sight. Faraday rotation is a powerful tool for measuring the magnetic field strength towards astrophysical objects. The combination of cosmic magnetic fields and charged particles distributed along the sightline towards a background radio source causes rotation of the polarization angle of linearly polarized synchrotron emission \citep[e.g.][]{2011hea..book.....L}. Along a line of sight, the observed polarization angle is altered by an amount equal to
\begin{equation}
\Phi = \Phi_{\textrm{0}} + \textrm{RM}\lambda^2 \,,
\end{equation}
where $\lambda$ is the observing wavelength, $\Phi$ and $\Phi_{\textrm{0}}$ are the measured and intrinsic polarization angles respectively, and the constant of proportionality RM, the `rotation measure', is generally related to the integrated product of the electron number density, \(n_{\rm e}\), and the strength of the component of the magnetic field parallel to the line of sight, \(B_{\parallel}\). The observed RM is also related to the redshift at which the Faraday rotating medium is located, but as it is generally not known where all of the rotating media are distributed along the line of sight, this relation is typically not simple. Nevertheless, measurements of the RM can be used to infer the presence of magnetic fields somewhere along the line of sight between an observer and a source.

Using a ``back-lit'' quasi-stellar object (QSO, which we will use interchangably with quasar) as a flashlight shining through foreground intervening material, it has previously been suggested that there is a correlation between metal-line absorption and the Faraday rotation/RM of distant polarized sources \citep[e.g.][]{1984ApJ...279...19W,1992ApJ...387..528K}. The experimental setup of a typical back-lit experiment is shown in Fig.~\ref{DLAcartoon}. Previous studies have already begun using the correlations seen in back-lit experiments to indirectly test dynamo theory in normal galaxies using strong magnesium~II (Mg\,{\sc ii}) absorption lines \citep[e.g.][]{2008Natur.454..302B,2013ApJ...772L..28B,2014ApJ...795...63F} located in the spectra of QSOs. This analysis can be extended to use Damped Lyman-Alpha Absorption systems and Lyman Limit Systems in the spectra of QSOs, and thereby probe the distribution of gaseous matter throughout the Universe, particularly in the protogalactic and the intergalactic medium \citep{1998ARA&A..36..267R,2005ARA&A..43..861W,2009RvMP...81.1405M}.

In this paper, we are therefore interested in two types of absorption systems, and define these systems via their neutral hydrogen column densities, with $1.6\times10^{17} <$ $N$(H\,{\sc i}) $< 2\times10^{20}$~cm$^{-2}$ being Lyman Limit Systems (LLSs), and $N$(H\,{\sc i}) $\ge 2\times10^{20}$~cm$^{-2}$ being Damped Lyman-Alpha Absorbers (DLAs) \citep{2005ARA&A..43..861W,2015MNRAS.451..904E}. There is a fundamental difference between both systems: hydrogen is mainly neutral in DLAs, while it is mainly ionised in LLSs. This characteristic separates the DLAs from both the LLSs and other intervening absorbers seen in QSO sightlines, such as the Ly-alpha forest ($N$(H\,{\sc i}) $\le10^{17}$~cm$^{-2}$), where the neutral gas is a minor or non-existent phase. The presence of neutral, cold, and molecular gas is crucial to link the DLAs to star-forming galaxies \citep{2003ApJ...593..215W,2003MNRAS.346..209L,2005ARA&A..43..861W,2005ApJ...622L..81H,2008A&A...481..327N}, although the exact nature of the DLAs and LLSs and their relation to present-day galaxies is still under debate and study \citep{2005ARA&A..43..861W,2006ApJ...652..981W,2008ApJ...681..856R,2009RvMP...81.1405M}. Although the nature of these absorption systems remains poorly understood, they possibly provide the only example of an interstellar medium in the high-redshift Universe.

Both types of system are important, as they are known to be some of the biggest intervening neutral-hydrogen gas reservoirs in the Universe and thereby constitute the building blocks of galaxies. The combination of both DLAs and LLSs provides a large range in column density that allows us to explore and contrast the difference between these relative ends of the absorption system extrema. Importantly, both DLAs and LLSs are generally believed to correspond to similar features in the intergalactic medium with both systems probing the progenitors of current massive galaxies \citep[e.g.][]{2010ApJ...721.1448S}, and with DLAs corresponding to extended disks \citep[e.g.][]{1998ApJ...507..113P} and LLSs corresponding to extended gaseous haloes \citep[e.g.][]{2015MNRAS.451..904E}. The column density of these absorbers has been found to be correlated to the mass of the nearest galaxy, with the correlation more pronounced for DLAs, and similar correlations found between the star-formation rate, halo mass, and H\,{\sc i} content of the associated galaxies \citep[e.g.][]{2014MNRAS.438..529R}. Similarly, LLSs display a correlation between $N$(H\,{\sc i}) and halo mass, with lower column density systems more likely to be found near lower mass halos \citep[e.g.][]{2012MNRAS.421.2809V}.

However, only two studies to-date have attempted to look for connections between protogalaxies themselves and magnetic fields. The first study of \citet{1992ApJ...388...17W} used Mg\,{\sc ii} systems with H\,{\sc i} column densities above $2\times10^{20}$~cm$^{-2}$, and classified these as DLAs. They found 5 background sources with Faraday rotation higher than 3 times the 1$\sigma$ uncertainty in RRM. The second study of \citet{1995ApJ...445..624O} used DLAs along the line of sight towards 11 background sources with known Faraday rotation. This data-limited study provided inconclusive results, with a ``tentative'' indication of higher Faraday rotation associated with the DLAs. Neither study attempted to measure the LLSs. This is a similar scenario to the previously weak correlations reported between |RM| and the number of strong Mg\,{\sc ii} absorption lines along the line of sight \citep{1982ApJ...263..518K,2012ApJ...761..144B}, which have more recently been expanded into definitive connections \citep{2014ApJ...795...63F}. The next step in understanding the magnetised structure, strength, and coherence in low-luminosity galaxies, is therefore to expand from studies of Mg\,{\sc ii} absorption lines to new studies of DLAs and LLSs. We can attempt to use polarization measurements to measure the magnetic fields in the ionized gas component of these systems \citep[e.g.][]{2015ApJ...808...38R}.

This paper is structured as follows: we present our observational data in Section~\ref{data}, which details and justifies how we created our sample and its properties. Our results, including a quantitative analysis of our main sample is detailed in Section~\ref{results}. A discussion of our results and the estimation of magnetic properties from our data is presented in Section~\ref{discussion}, while our conclusions are summarised in Section~\ref{conclusions}. We refer to `polarization' on multiple occasions, in all cases we are referring to linear radio polarization. Unless otherwise specified, all quantities are presented as measured in the observed-frame.

\section{Sample Construction}
\label{data}
To create the sample, we have concatenated various DLA and LLS identifications taken from across the literature. As our initial starting catalogue, we have used the ninth data release of the Sloan Digital Sky Survey (SDSS) Quasar Catalog, DR9Q, of 87,822 sources \citep{2012A&A...548A..66P}. We have cross-matched this catalogue to the original NRAO Very Large Array Sky Survey (NVSS) \citep{1998AJ....115.1693C}. In order to only incorporate those sources with the most definitive, single-component, compact optical and radio counterparts, we use a 2~arcsec cross-matching radius. This yields 1,253 sources, and is our initial control sample. We have then used a DLA catalogue of 12,081 quasars that were observed in the course of the Baryon Oscillation Spectroscopic Survey (BOSS), which is also part of the SDSS DR9 \citep{2012A&A...547L...1N}. Each of the quasars has one detected intervening DLA or LLS along the line-of-sight. We have also cross-matched this catalogue to the original NVSS with a 2~arcsec radius. This yields 164 sources, and is our initial sample of absorbers.

Both the radio-matched control catalogue of 1,253 sources \citep{2012A&A...548A..66P}, and also the radio-matched DLA/LLS catalogue of 164 sources \citep{2012A&A...547L...1N}, were derived using the same dataset. In \citet{2012A&A...547L...1N}, all sources from \citet{2012A&A...548A..66P} were searched for DLAs/LLSs, and only those with positive detections were listed. Further details of the criteria used to identify DLAs/LLSs are provided in \citep{2009A&A...505.1087N,2012A&A...547L...1N}. Consequently, we therefore remove sources from the control sample in which DLAs/LLSs were positively identified. This reduces the control sample to 1,090 sources in which no intervening DLAs or LLSs were identified. We then proceed to incorporate various other DLA catalogues gathered from throughout the literature, and that we again pre-match with the original NVSS using a 2~arcsec radius. The sample of \citet{2009A&A...508..133G} provides two sources, \citet{2013MNRAS.435..482J} provides five sources, \citet{2006ApJ...636..610R} provides 43 sources, \citet{2014A&A...566A..24N} provides two sources, \citet{2010ApJ...718..392P} provides 50 sources, and \citet{2002PASA...19..455C} provides 39 sources. We also attempted to include data from \citet{2015MNRAS.449.1536T}, but identified no NVSS radio counterparts.
\begin{figure}
\centering
\includegraphics[trim=0.5cm 0.5cm 0.5cm 0.0cm, clip=true, width=\hsize]{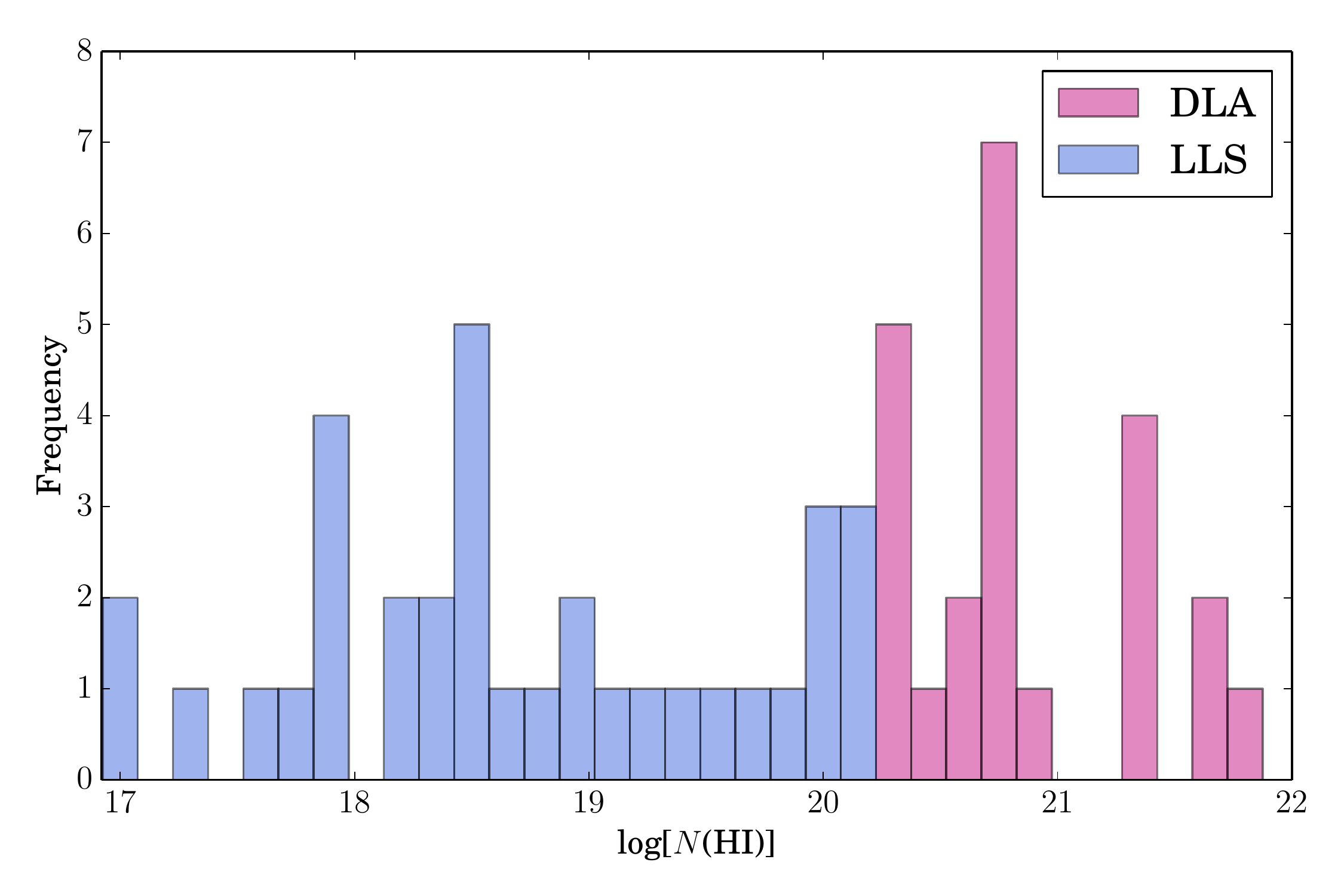}
\caption{The column density distribution for our samples. Lines-of-sight categorised into the DLA sample (violet), and the LLS sample (blue) are shown in the histograms.}
\label{NHI}
\end{figure}

In order to check whether there were any duplicated matches, we cross-checked these additional data against the initially formed control and DLA/LLS samples. Of the five sources from \citet{2013MNRAS.435..482J}, two sources were found to have existing matches in the DLA/LLS sample, with manual inspection showing consistent estimates for the redshift of the absorber and the neutral hydrogen column density. As the data of \citet{2013MNRAS.435..482J} were derived from SDSS DR5, we instead propagated the DR9 match into the final absorber catalogue. The additional three sources without existing records were also added into the sample of absorbers. Similarly, of the two sources from \citet{2014A&A...566A..24N}, a single source was found to have an existing match in the DLA/LLS sample. As the sources from \citet{2014A&A...566A..24N} are from SDSS DR11, the existing record for this one datapoint was replaced. However, manual inspection again showed that the redshift of the absorber and the neutral hydrogen column density of both data were in agreement within their uncertainties. The additional one source without an existing record was added into the sample. Finally, of the 39 sources from \citet{2002PASA...19..455C}, a single source was found to have an existing match in the control sample and a further nine sources were found to have existing matches in the DLA/LLS sample. As \citet{2002PASA...19..455C} use older data, we rely on the accuracy of the \citet{2012A&A...548A..66P} data, and discard the data of the single source with a \citet{2002PASA...19..455C} match. For the nine sources in the DLA/LLS sample, manual inspection showed that for seven sources the redshift of the absorber and the neutral hydrogen column density of both data were in agreement within their uncertainties. Nevertheless, two sources have updated estimates of source parameters in the contemporary data -- implying either a $\sim7/9=77$\% accuracy rate in the earlier measurements, or possibly a systematic affecting the newer data. However, both the old and new data are in agreement that these sightlines do contain intervening DLAs/LLSs. We therefore keep our current estimates, and do not update our catalogue values using these nine sources. The remaining 29 sources are added into the DLA/LLS sample -- with an awareness that $\approx6$ of these sources may not have accurate estimates of the source parameters. For some sources, Ww cannot rule out e.g.\ low-redshift intervenors observable in the ultraviolet part of the spectrum, due to the spectral-coverage provided. However, there is no reason to expect any differences between the control, DLA, and LLS samples. Effectively all of our samples should contain low-redshift intervenors to a similar degree, and so this cannot explain any detected signal.

\begin{figure}
\centering
\includegraphics[trim=0cm 0.0cm 0cm 0.0cm, clip=true, width=\hsize]{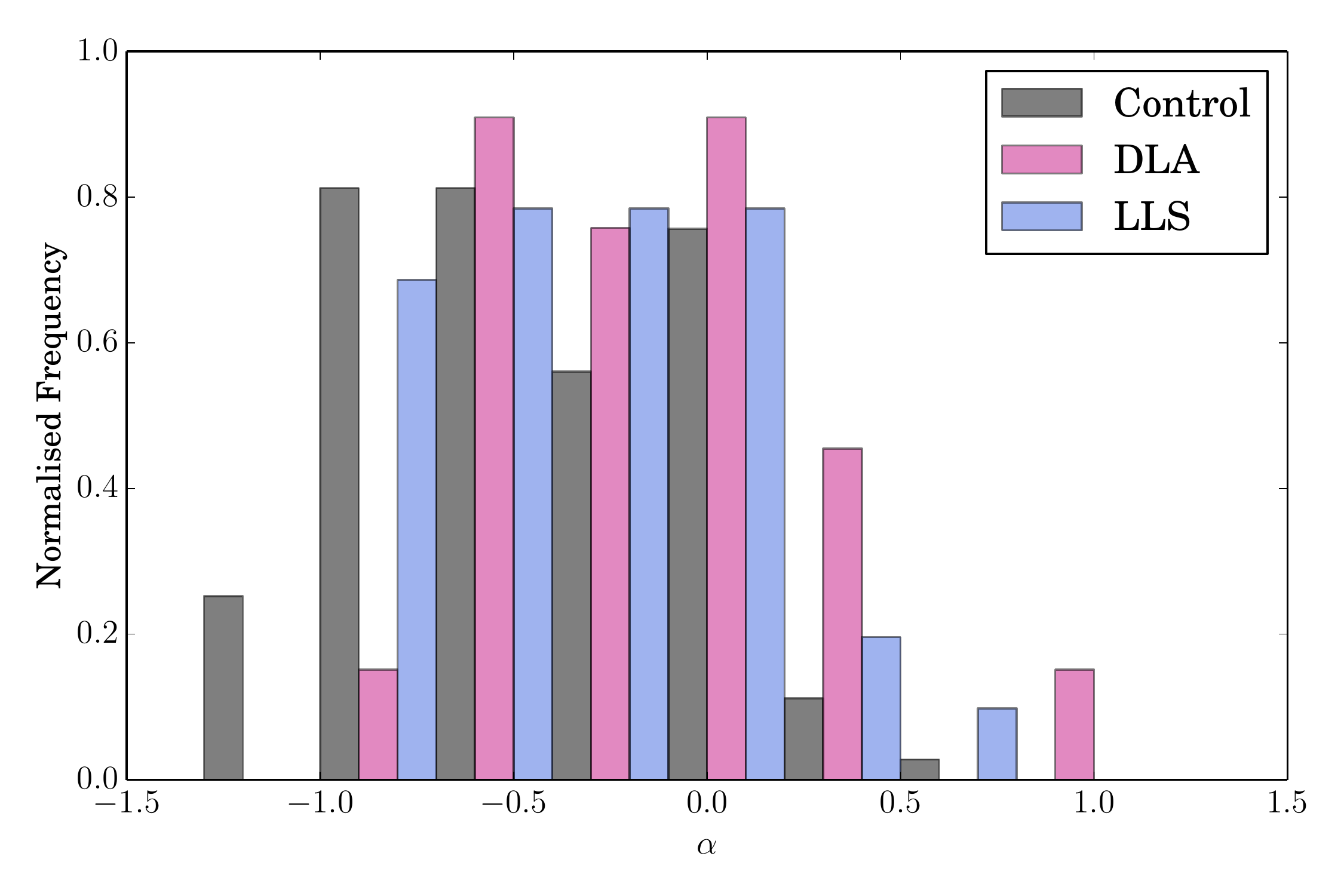}
\caption{The spectral index distribution for our sample. Lines-of-sight categorised into the control sample (black), the DLA sample (violet), and the LLS sample (blue) are all shown in the normalised histogram.}
\label{spectra}
\end{figure}

We have created a control sample of 1,090 sources with no known DLAs/LLSs, and a sample of 242 sources with an intervening DLA/LLS along the line-of-sight. From cross-matching the catalogue of \citet{2002PASA...19..455C}, we found that $\approx6$ sources may not have accurate source parameters, indicating an upper limit on the false-detection rate (FDR) of $\le6/242=2.5$\%. Nevertheless, as this only affects our oldest catalogue, the FDR is likely considerably lower. Each quasar has both an optical detection and a 1.4~GHz radio counterpart. As all data were cross-matched using a very tight 2~arcsec cross-matching radius, we have only included the most definitive matches into our sample. This should alleviate the need to split our sample by spectral index as a proxy for enhanced resolution, as has been done in studies of strong Mg\,{\sc ii} absorption \citep{2014ApJ...795...63F}. Unlike strong Mg\,{\sc ii}, which is associated with a clumpy partially-ionised medium, DLAs and LLSs are associated with a smooth medium with high covering fraction (see Section~\ref{discussion}). The spectral index criterion is therefore likely not an essential requirement for DLAs/LLSs (although we lack the sample size to thoroughly test this hypothesis). The spectral index distribution of the final sample of sources is shown in Fig.~\ref{spectra}.

\begin{figure}
\centering
\includegraphics[trim=0cm 0.0cm 0cm 0.0cm, clip=true, width=\hsize]{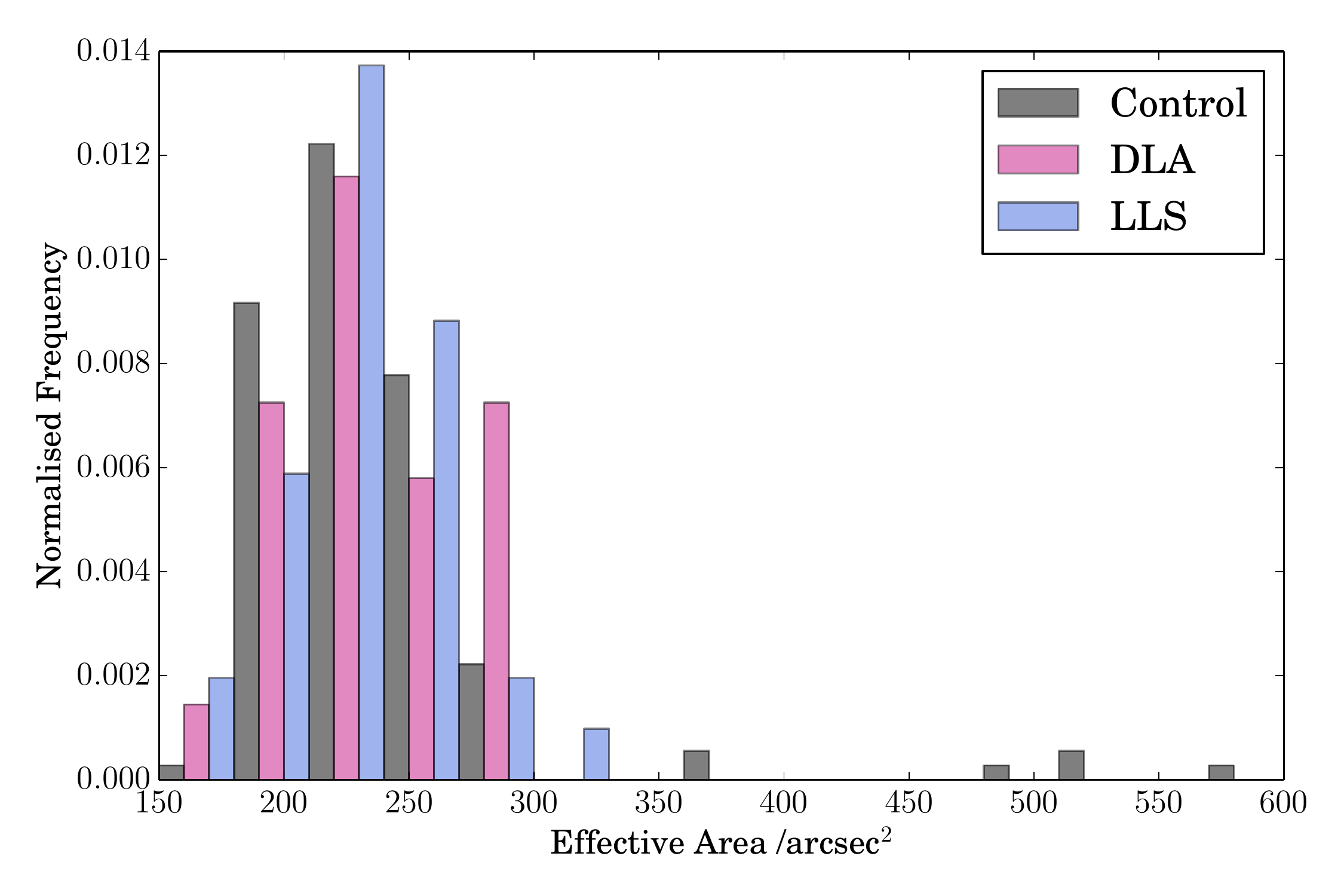}
\caption{The distribution of the effective area for each source in our sample. Lines-of-sight categorised into the control sample (black), the DLA sample (violet), and the LLS sample (blue) are all shown in the normalised histogram.}
\label{areas}
\end{figure}

As we are here chiefly interested in the radio polarization, we continue to cross-match our samples with the NVSS catalogue of \citet{2009ApJ...702.1230T}, in which all sources are polarized at $\ge8\sigma$. This high signal--to--noise (s/n) detection threshold is crucial so that Rician bias effects are negligible \citep{1985A&A...142..100S}. If Rician bias effects were significant, then our experimental design would be fundamentally flawed. Fainter sources have larger bias, and so these sources would appear more polarized than they truly are. To counter this effect, we would need to check that all of the samples cover \emph{similar distributions} of polarized and total intensity. However, this would not be possible, as one of our primary science questions is whether the samples have different polarized fractions -- which depends on \emph{different distributions} of polarized and total intensity. Ultimately, this would result in a circular argument. For the cross-matching, we use an arbitrary 2~arcsec cross-matching radius to combine the two versions of the NVSS. We also use the original NVSS catalogue in order to calculate the effective area of each source, using $A = \frac{1}{4} \pi \theta_{M} \theta_{m}$, where $\theta_{M/m}$ are the major and minor axes respectively. The effective area distribution of the final sample of sources is shown in Fig.~\ref{areas}.

\begin{figure}
\centering
\includegraphics[trim=0cm 0.0cm 0cm 0.0cm, clip=true, width=\hsize]{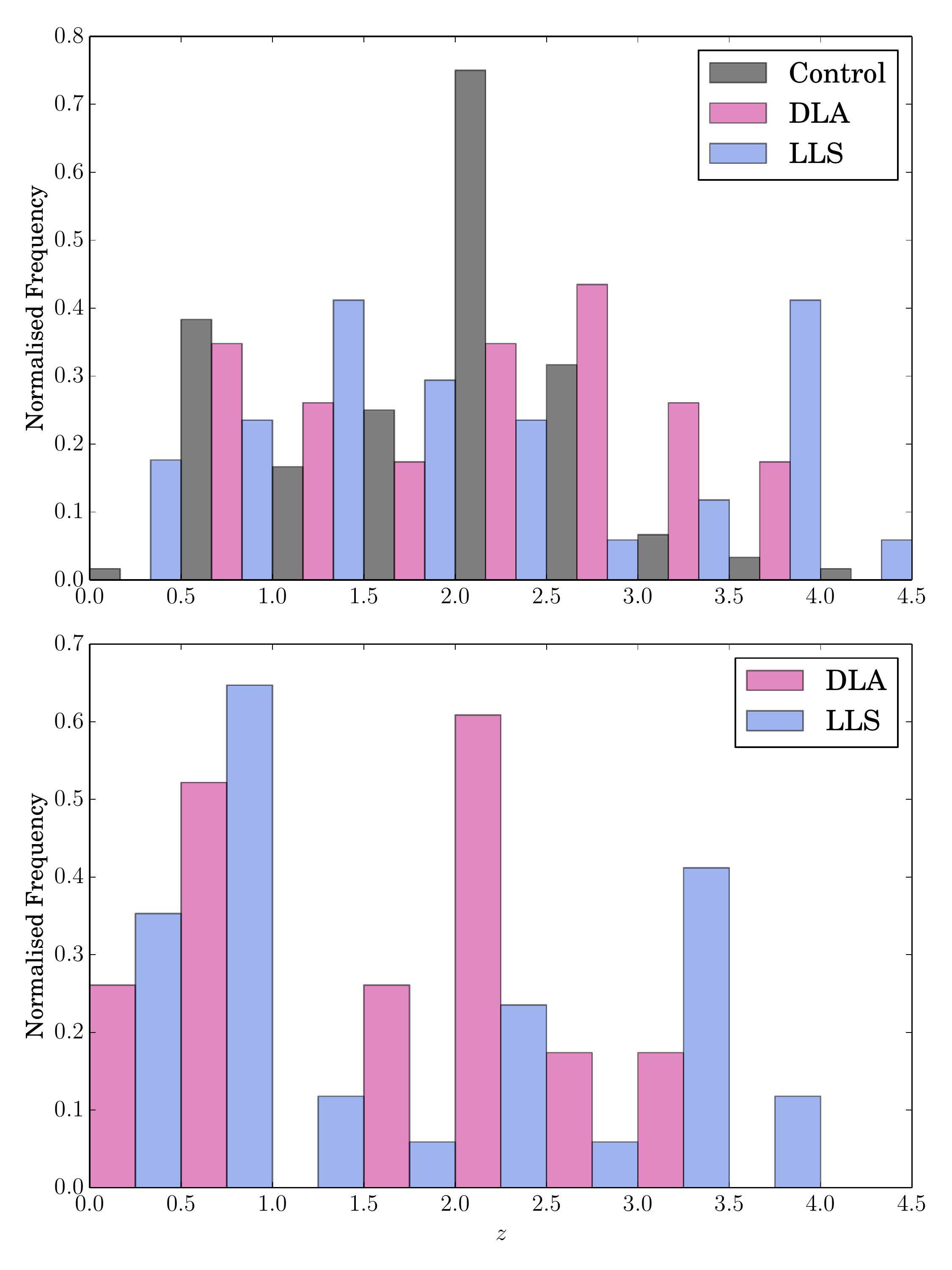}
\caption{The redshift distributions for our sample. Lines-of-sight categorised into the control sample (black), the DLA sample (violet), and the LLS sample (blue) are all shown in the normalised histograms. \textbf{Top:} The redshift distribution of the background QSOs. \textbf{Bottom:} The redshift distribution of the absorbing systems.}
\label{redshifts}
\end{figure}

Some of our catalogues have varying and inconsistent definitions of the column densities associated with DLAs and LLSs, and so we re-sort the data such that systems with $1.6\times10^{17} <$ $N$(H\,{\sc i}) $< 2\times10^{20}$~cm$^{-2}$ are classified as LLSs, and $N$(H\,{\sc i}) $\ge 2\times10^{20}$~cm$^{-2}$ are classified as DLAs \citep[e.g.][]{2015MNRAS.451..904E}. The column density distribution of the final sample of sources is shown in Fig.~\ref{NHI}. This yields a control sample of 120 sources, a sample of 23 DLAs, and 34 LLSs -- a total of 177 sources. All of these sources have known radio polarization fractions and RMs. Note that this updated sample has 5 times more Lyman-absorbing systems than the largest published study in this area \citep{1995ApJ...445..624O}, which even then did not include any LLSs. Our sample includes ultraviolet and optically-identified DLAs/LLSs -- and covers a significant range in redshift-space with absorbers from $0.23\le z \le 3.81$. The redshift distribution of the final sample of sources is shown in Fig.~\ref{redshifts}.

In order to remove the effects of the Galactic foreground, we also excluded sources located at Galactic latitudes, $|b|\le25^{\circ}$. This leaves a total of 114 control, 19 DLA, and 27 LLS sources. The properties of each source used in our analysis are given in Tables~\ref{table:1} to ~\ref{table:1b}.
\begin{deluxetable*}{c c c c c c c c c c}
\tablewidth{0pt}
\tablecaption{Source parameters for all of the polarized QSOs used in our analysis. All of the source names, uncertainties, and more extensive source parameters can be found in \citet{2009ApJ...702.1230T}, \citet{2014ApJS..212...15F}, and \citet{2015A&A...575A.118O}. The control, DLA, and LLS samples are denoted ``C'', ``D'', and ``L'' respectively. Blanks entries have no listed value. Continued in Tables~\ref{table:1a} and \ref{table:1b}.}
\tablehead{
\colhead{R.A.} & \colhead{Dec.}  & \colhead{RM}  & \colhead{GRM} & \colhead{$\Pi$} & \colhead{$\alpha$} & \colhead{$z_{\textrm{qso}}$} & \colhead{$z_{\textrm{abs}}$} & \colhead{log[$N$(H\,{\sc i}) /cm$^{-2}$]}  & \colhead{Type}  \\
\colhead{(degrees)} & \colhead{(degrees)}  & \colhead{(rad~m$^{-2}$)}  & \colhead{(rad~m$^{-2}$)} & \colhead{($\%$)} & \colhead{} & \colhead{} & \colhead{} & \colhead{} & \colhead{(C/D/L)} }
\startdata
5.901 & -2.885 & -29.1 & -3.4 & 4.05 & -0.70 & 2.74 &     &     & C \\
16.596 & -1.927 & -9.5 & 1.6 & 7.62 & -0.33 & 2.19 &     &     & C \\
17.112 & -0.623 & 6.2 & 6.6 & 6.91 & -0.31 & 1.38 &     &     & C \\
22.752 & 0.290 & 15.5 & 13.5 & 8.95 & -1.07 & 4.02 &     &     & C \\
23.652 & 6.210 & -1.7 & 3.3 & 4.32 & -0.96 & 1.00 &     &     & C \\
25.080 & 14.774 & -12.2 & -9.6 & 11.82 & -0.47 & 2.23 &     &     & C \\
27.343 & 5.931 & 28.8 & 9.4 & 0.91 & 0.26 & 2.35 &     &     & C \\
115.482 & 42.138 & -0.9 & 1.1 & 5.02 & -0.65 & 2.23 &     &     & C \\
115.656 & 39.743 & 0.3 & 2.6 & 3.15 & -0.15 & 2.20 &     &     & C \\
117.122 & 50.888 & -9.4 & -1.7 & 1.82 & -0.95 & 0.63 &     &     & C \\
120.067 & 53.672 & -47.0 & -8.0 & 12.34 & -0.56 & 1.93 &     &     & C \\
120.228 & 45.014 & 9.9 & 11.2 & 9.12 & -0.42 & 2.36 &     &     & C \\
121.243 & 49.813 & -6.6 & -3.3 & 6.73 & -0.57 & 2.96 &     &     & C \\
121.640 & 45.075 & 23.9 & 20.7 & 2.24 & 0.14 & 2.11 &     &     & C \\
121.954 & 45.264 & 33.3 & 21.3 & 4.65 & -0.52 & 3.19 &     &     & C \\
122.905 & 40.789 & 17.5 & 22.7 & 5.85 & -0.79 & 0.67 &     &     & C \\
123.448 & 53.798 & -19.9 & -6.4 & 4.24 & -0.52 & 0.95 &     &     & C \\
123.599 & 19.984 & -15.3 & 4.7 & 5.61 & -0.86 & 1.50 &     &     & C \\
123.893 & 33.091 & 2.6 & 5.9 & 2.02 & -0.88 & 2.42 &     &     & C \\
124.263 & 16.383 & 42.5 & 38.4 & 4.59 & -1.00 & 0.70 &     &     & C \\
125.921 & 29.475 & 9.6 & 11.9 & 2.02 & -0.03 & 2.37 &     &     & C \\
126.125 & 31.726 & -3.7 & 25.2 & 9.96 & -1.11 & 2.37 &     &     & C \\
126.515 & 52.795 & -29.3 & -10.0 & 5.15 & -0.41 & 1.53 &     &     & C \\
126.519 & 36.115 & 26.2 & 20.3 & 3.16 & -0.69 & 1.03 &     &     & C \\
126.778 & 10.873 & 19.1 & 20.6 & 6.38 & -0.16 & 2.28 &     &     & C \\
127.039 & 29.339 & 18.2 & 19.7 & 1.76 & -0.51 & 2.32 &     &     & C \\
128.206 & 15.902 & 14.3 & 15.3 & 1.27 & -0.06 & 2.42 &     &     & C \\
128.310 & 11.393 & 52.8 & 40.9 & 1.49 & -0.26 & 2.98 &     &     & C \\
129.095 & 27.481 & 24.0 & 17.4 & 1.67 & -0.10 & 0.76 &     &     & C \\
129.205 & 11.784 & 26.9 & 22.2 & 4.8 &  & 2.41 &     &     & C \\
131.121 & 38.515 & 9.2 & 15.2 & 1.02 & -0.35 & 2.94 &     &     & C \\
136.192 & 2.145 & -3.6 & -4.4 & 4.12 & -0.65 & 0.79 &     &     & C \\
137.292 & 1.360 & -18.0 & -14.4 & 6.22 & 0.17 & 1.02 &     &     & C \\
137.316 & 3.912 & 38.0 & 35.6 & 5.4 & 0.02 & 3.29 &     &     & C \\
137.920 & 2.088 & -12.2 & -2.4 & 3.25 & -0.01 & 2.37 &     &     & C \\
143.811 & 1.537 & -38.0 & -16.4 & 4.78 & -0.80 & 2.55 &     &     & C \\
143.882 & 36.555 & -5.3 & 14.3 & 1.5 & 0.11 & 2.85 &     &     & C \\
149.512 & 38.500 & -33.8 & 7.5 & 1.81 & -0.97 & 1.40 &     &     & C \\
154.546 & 35.711 & 6.4 & 7.1 & 2.88 & 0.07 & 1.23 &     &     & C \\
154.920 & 4.733 & -7.0 & 10.1 & 10.83 & -0.23 & 2.29 &     &     & C \\
156.249 & 6.415 & 68.7 & 21.3 & 1.45 & -1.01 & 1.74 &     &     & C \\
156.633 & 6.459 & 29.7 & 21.3 & 1.6 & -0.86 & 2.46 &     &     & C \\
157.120 & 35.885 & -12.5 & 5.6 & 3.51 & -0.42 & 2.60 &     &     & C \\
160.175 & 36.903 & 22.5 & 11.0 & 8.24 & -0.68 & 2.69 &     &     & C \\
161.044 & 35.153 & 1.0 & 16.9 & 4.9 & -0.88 & 2.22 &     &     & C \\
163.632 & 38.923 & 18.5 & 9.5 & 7.51 & -0.16 & 1.37 &     &     & C \\
165.932 & 41.219 & 28.8 & 12.2 & 2.07 & -0.61 & 2.46 &     &     & C \\
168.161 & 34.778 & 48.5 & 10.9 & 5.69 & 0.01 & 1.95 &     &     & C \\
171.170 & 2.589 & 26.2 & 5.3 & 3.18 & -0.84 & 0.78 &     &     & C \\
174.712 & 4.478 & 17.5 & 9.7 & 5.26 & -0.03 & 2.38 &     &     & C \\
174.761 & 40.549 & 6.4 & 5.5 & 5.29 & -0.11 & 2.36 &     &     & C \\
183.954 & 31.860 & -0.6 & 1.4 & 4.13 & -0.17 & 2.26 &     &     & C \\
184.145 & 37.391 & -19.8 & -6.4 & 1.78 & -0.87 & 0.83 &     &     & C \\
188.053 & 33.928 & -5.2 & -3.9 & 6.0 & -0.88 & 2.48 &     &     & C \\
192.636 & 2.276 & -26.9 & -8.5 & 3.05 & -0.03 & 3.50 &     &     & C \\
193.189 & 5.284 & -22.9 & 2.2 & 2.96 & -0.78 & 0.63 &     &     & C \\
193.593 & 5.772 & -3.1 & 3.7 & 5.72 & -0.24 & 2.25 &     &     & C \\
196.183 & 1.542 & -16.3 & -10.4 & 11.19 & -0.14 & 2.29 &     &     & C \\
200.482 & 35.177 & 19.6 & 2.8 & 3.33 & -0.59 & 1.92 &     &     & C \\
204.568 & 39.475 & 13.1 & 4.5 & 4.09 & -0.52 & 2.46 &     &     & C \\
206.403 & 38.387 & 16.7 & 5.3 & 0.54 & -0.65 & 1.85 &     &     & C \\
208.298 & 32.095 & -10.3 & 0.3 & 1.89 & -0.69 & 1.56 &     &     & C \\
209.863 & 1.998 & 11.4 & 2.6 & 5.86 & 0.05 & 1.33 &     &     & C \\
210.830 & 35.137 & -12.9 & 2.2 & 5.32 & -0.99 & 2.29 &     &     & C \\
211.191 & -1.506 & -24.1 & -5.4 & 3.62 & 0.15 & 2.52 &     &     & C \\
213.869 & 37.106 & 24.4 & 2.9 & 1.56 & 0.21 & 2.37 &     &     & C \\
214.779 & 5.918 & -11.4 & -0.1 & 4.04 & -0.41 & 2.29 &     &     & C \\
215.480 & 3.927 & -9.2 & -0.2 & 2.82 & -0.40 & 1.00 &     &     & C \\
223.327 & 35.094 & 29.6 & 4.6 & 2.17 & -0.06 & 0.71 &     &     & C \\
224.409 & 7.832 & -3.7 & 10.7 & 4.68 & 0.73 & 0.57 &     &     & C \\
224.489 & 34.664 & 5.5 & 5.8 & 2.3 & -0.39 & 2.73 &     &     & C \\
226.111 & 28.909 & -1.9 & 3.2 & 2.72 & -0.63 & 2.28 &     &     & C \\
226.305 & 6.909 & 25.8 & 16.0 & 5.07 & -0.63 & 2.71 &     &     & C \\
230.855 & 27.083 & 31.0 & 10.2 & 4.88 & -1.03 & 2.19 &     &     & C 
\enddata
\label{table:1}
\end{deluxetable*}

\begin{deluxetable*}{c c c c c c c c c c}
\tablewidth{0pt}
\tablecaption{Continued from Table~\ref{table:1}. Source parameters for all of the polarized QSOs used in our analysis. The table is continued further in Table~\ref{table:1b}.}
\tablehead{
\colhead{R.A.} & \colhead{Dec.}  & \colhead{RM}  & \colhead{GRM} & \colhead{$\Pi$} & \colhead{$\alpha$} & \colhead{$z_{\textrm{qso}}$} & \colhead{$z_{\textrm{abs}}$} & \colhead{log[$N$(H\,{\sc i}) /cm$^{-2}$]}  & \colhead{Type}  \\
\colhead{(degrees)} & \colhead{(degrees)}  & \colhead{(rad~m$^{-2}$)}  & \colhead{(rad~m$^{-2}$)} & \colhead{($\%$)} & \colhead{} & \colhead{} & \colhead{} & \colhead{} & \colhead{(C/D/L)} }
\startdata
232.255 & -1.090 & 11.9 & 8.5 & 5.63 & -1.25 & 2.24 &     &     & C \\
233.313 & 13.540 & 20.0 & 18.9 & 3.36 & -0.79 & 0.77 &     &     & C \\
234.855 & 16.067 & 12.7 & 16.6 & 7.83 & -0.06 & 2.53 &     &     & C \\
234.913 & 27.744 & 6.5 & 12.2 & 2.82 & 0.05 & 2.20 &     &     & C \\
236.248 & 4.130 & 22.1 & 12.0 & 2.35 & -0.59 & 2.18 &     &     & C \\
238.912 & 11.112 & 29.9 & 13.2 & 5.0 & -0.09 & 2.66 &     &     & C \\
239.415 & 22.644 & -2.6 & 20.2 & 8.39 & -0.53 & 0.72 &     &     & C \\
239.820 & 11.263 & -6.7 & 7.4 & 0.94 & -0.56 & 1.94 &     &     & C \\
239.879 & 3.080 & 3.2 & 4.3 & 1.6 & 0.08 & 3.89 &     &     & C \\
240.011 & 4.216 & 10.4 & 7.9 & 10.01 & -0.19 & 0.79 &     &     & C \\
240.071 & 18.642 & 18.3 & 30.9 & 3.42 & -0.21 & 2.40 &     &     & C \\
240.553 & 24.170 & 26.4 & 22.1 & 1.42 & -0.94 & 2.53 &     &     & C \\
241.388 & 30.025 & 17.6 & 10.7 & 4.36 & 0.12 & 2.41 &     &     & C \\
241.615 & 31.435 & 18.3 & 8.9 & 1.64 & -0.95 & 1.94 &     &     & C \\
242.352 & 6.192 & 18.2 & 19.7 & 5.2 & -0.74 & 0.79 &     &     & C \\
244.232 & 36.360 & 14.0 & 8.6 & 2.0 & -0.10 & 2.26 &     &     & C \\
246.303 & 29.556 & -5.2 & 17.8 & 2.46 & -1.12 & 1.54 &     &     & C \\
246.894 & 20.013 & 48.5 & 43.1 & 4.88 & -1.10 & 1.53 &     &     & C \\
247.786 & 20.380 & 34.6 & 41.7 & 3.47 & -0.86 & 0.81 &     &     & C \\
249.159 & 21.215 & 54.0 & 42.7 & 1.15 & 0.04 & 1.80 &     &     & C \\
249.566 & 27.934 & 20.3 & 29.4 & 3.89 & -0.71 & 2.18 &     &     & C \\
251.219 & 18.222 & 13.2 & 29.8 & 3.66 & 0.13 & 0.79 &     &     & C \\
252.612 & 34.926 & 17.0 & 19.4 & 4.85 & -0.72 & 0.19 &     &     & C \\
253.932 & 19.813 & 48.0 & 51.7 & 3.78 & 0.15 & 3.26 &     &     & C \\
254.506 & 34.724 & 89.3 & 20.9 & 1.96 & -0.01 & 1.94 &     &     & C \\
257.479 & 22.616 & 86.3 & 67.9 & 1.33 & -1.02 & 1.54 &     &     & C \\
260.721 & 24.976 & 45.8 & 51.1 & 3.0 & -0.35 & 2.25 &     &     & C \\
261.958 & 34.378 & 38.4 & 40.2 & 6.6 & -0.96 & 2.43 &     &     & C \\
262.566 & 35.210 & 76.5 & 51.6 & 5.81 & -0.76 & 0.58 &     &     & C \\
318.196 & -1.568 & -18.4 & -10.6 & 2.72 & -0.89 & 0.77 &     &     & C \\
319.153 & 5.605 & 33.9 & 30.9 & 6.63 & 0.25 & 2.22 &     &     & C \\
320.111 & 4.802 & -2.3 & 8.5 & 4.58 & -0.26 & 3.15 &     &     & C \\
326.980 & 8.503 & -24.2 & -21.2 & 4.64 & -0.30 & 2.60 &     &     & C \\
329.304 & 10.240 & -22.6 & -14.6 & 1.56 & 0.13 & 0.76 &     &     & C \\
333.789 & 13.377 & 21.8 & -30.4 & 4.94 & -0.97 & 1.90 &     &     & C \\
336.694 & 0.870 & -16.0 & -5.7 & 0.95 & -0.19 & 2.26 &     &     & C \\
344.741 & 2.061 & -56.9 & -19.3 & 2.43 & -0.33 & 2.67 &     &     & C \\
349.030 & 1.004 & -45.1 & -23.3 & 4.74 & -0.57 & 2.63 &     &     & C \\
357.578 & -0.116 & 20.2 & -5.0 & 1.86 & -0.86 & 1.37 &     &     & C \\
359.620 & 4.507 & -52.9 & -6.5 & 1.37 & 0.24 & 2.30 &     &     & C \\
33.654 & 6.548 & 3.4 & -0.9 & 3.86 & -0.37 & 2.31 & 2.11 & 20.8 & D \\
140.149 & 0.392 & 1.1 & -3.4 & 4.66 & -0.40 & 2.48 & 2.04 & 20.6 & D \\
232.466 & 19.076 & -29.0 & 14.7 & 2.55 & -0.56 & 2.36 & 2.22 & 20.7 & D \\
239.654 & 24.637 & 26.8 & 20.5 & 15.67 &  & 3.35 & 2.20 & 20.3 & D \\
332.734 & 1.986 & -6.7 & -14.3 & 5.3 & -0.17 & 2.59 & 2.39 & 20.5 & D \\
127.717 & 24.183 & 12.0 & 15.4 & 8.75 & 0.00 & 0.94 & 0.52 & 20.3 & D \\
148.737 & 17.725 & -5.5 & 2.5 & 4.42 & -0.24 & 1.48 & 0.24 & 21.3 & D \\
172.529 & -14.824 & 34.0 & -14.0 & 4.44 & -0.05 & 1.19 & 0.31 & 21.7 & D \\
247.939 & 11.934 & 39.9 & 34.3 & 2.37 & -0.52 & 1.79 & 0.53 & 20.7 & D \\
20.132 & -27.024 & 18.6 & 12.3 & 4.03 & -0.10 & 0.56 & 0.56 & 20.3 & D \\
30.944 & 11.579 & -3.8 & -11.5 & 1.54 & 0.01 & 3.61 & 3.39 & 21.3 & D \\
39.662 & 16.616 & 42.0 & -0.7 & 2.02 & 0.00 & 0.94 & 0.52 & 21.6 & D \\
54.754 & -1.555 & 26.9 & 22.7 & 4.71 & -0.10 & 3.20 & 3.06 & 21.2 & D \\
57.491 & -21.047 & 26.2 & 33.0 & 1.44 & 0.42 & 2.94 & 1.95 & 20.3 & D \\
61.891 & -33.063 & 4.2 & 8.2 & 3.09 & -0.03 & 2.57 & 2.57 & 20.6 & D \\
82.533 & -25.058 & 32.2 & 24.8 & 0.68 & -0.41 & 2.78 & 2.14 & 20.6 & D \\
138.965 & 0.120 & 61.3 & -0.4 & 4.78 & -0.50 & 3.07 & 2.77 & 20.3 & D \\
187.732 & -11.653 & -36.2 & -11.8 & 0.98 & -0.57 & 3.53 & 2.19 & 20.6 & D \\
202.785 & 30.509 & 8.8 & 1.3 & 9.16 & -0.57 & 0.85 & 0.69 & 21.3 & D \\
159.414 & 4.400 & 76.8 & 11.5 & 2.53 & 0.12 & 2.32 & 2.15 & 20.2 & L \\
190.541 & 37.335 & -14.6 & -7.0 & 2.46 & -0.13 & 3.82 & 3.01 & 20.0 & L \\
224.863 & 32.900 & 28.1 & 7.7 & 4.7 & -0.91 & 3.33 & 2.23 & 20.1 & L \\
251.553 & 15.617 & 47.5 & 32.2 & 6.1 & -0.23 & 2.86 & 2.77 & 20.3 & L \\
20.616 & -4.358 & 27.7 & 0.1 & 2.55 & -0.67 & 1.95 & 0.66 & 18.8 & L \\
34.454 & 1.747 & -17.5 & -9.6 & 1.41 & 0.43 & 1.72 & 1.34 & 19.9 & L \\
65.816 & -1.342 & -25.5 & -22.9 & 1.67 & 0.52 & 0.92 & 0.63 & 18.5 & L \\
131.447 & 13.483 & 10.6 & 13.8 & 4.49 & -0.61 & 1.88 & 0.61 & 19.6 & L \\
160.321 & 6.171 & 25.0 & 15.2 & 2.17 & 0.09 & 1.26 & 0.44 & 18.3 & L \\
160.686 & 12.059 & 25.5 & 14.7 & 6.88 & -0.51 & 1.03 & 0.66 & 18.4 & L \\
163.136 & 61.422 & -1.6 & -0.8 & 3.9 & -0.80 & 0.42 & 0.23 & 18.0 & L \\
177.872 & 38.431 & 8.4 & 1.8 & 4.02 & -0.75 & 1.30 & 0.55 & 18.0 & L \\
187.153 & 10.312 & 2.6 & 2.5 & 3.24 & -0.80 & 2.31 & 0.94 & 19.4 & L \\
191.045 & 17.351 & -1.0 & 1.4 & 1.38 & -0.35 & 1.28 & 0.55 & 18.9 & L \\
201.373 & 65.254 & 40.2 & 33.3 & 3.31 & -0.88 & 1.62 & 1.52 & 18.6 & L \\
209.268 & 19.319 & 28.1 & 7.9 & 6.14 & 0.02 & 0.72 & 0.46 & 18.5 & L 
\enddata
\label{table:1a}
\end{deluxetable*}

\begin{deluxetable*}{c c c c c c c c c c}
\tablewidth{0pt}
\tablecaption{Continued from Table~\ref{table:1a}. Source parameters for all of the polarized QSOs used in our analysis.}
\tablehead{
\colhead{R.A.} & \colhead{Dec.}  & \colhead{RM}  & \colhead{GRM} & \colhead{$\Pi$} & \colhead{$\alpha$} & \colhead{$z_{\textrm{qso}}$} & \colhead{$z_{\textrm{abs}}$} & \colhead{log[$N$(H\,{\sc i}) /cm$^{-2}$]}  & \colhead{Type}  \\
\colhead{(degrees)} & \colhead{(degrees)}  & \colhead{(rad~m$^{-2}$)}  & \colhead{(rad~m$^{-2}$)} & \colhead{($\%$)} & \colhead{} & \colhead{} & \colhead{} & \colhead{} & \colhead{(C/D/L)} }
\startdata
322.897 & -12.118 & 13.0 & 9.9 & 1.93 & 0.14 & 0.50 & 0.43 & 19.2 & L \\
336.447 & -4.951 & -22.9 & -11.8 & 4.81 & -0.09 & 1.40 & 0.85 & 18.5 & L \\
323.550 & -4.319 & -7.4 & -7.2 & 4.83 & -0.49 & 4.33 & 3.27 & 20.0 & L \\
324.755 & 14.393 & -40.6 & -34.0 & 0.87 & 0.10 & 2.43 & 2.13 & 19.8 & L \\
118.264 & 42.525 & 147.0 & 18.1 & 1.25 & -0.14 & 3.59 & 3.22 & 18.6 & L \\
129.981 & 2.863 & -3.0 & 32.5 & 5.07 & -0.24 & 3.68 & 3.57 & 17.4 & L \\
133.238 & 24.517 & 13.1 & 21.5 & 1.8 & -0.73 & 3.62 & 3.34 & 17.0 & L \\
190.541 & 37.335 & -14.6 & -7.0 & 2.46 & -0.13 & 3.84 & 3.41 & 17.6 & L \\
208.359 & 57.431 & 55.7 & 15.0 & 2.68 & -0.66 & 3.47 & 3.46 & 17.8 & L \\
208.529 & -2.101 & -7.5 & -2.2 & 0.83 & -0.01 & 3.72 & 3.45 & 18.2 & L \\
218.891 & 54.600 & 14.6 & 12.7 & 4.26 & -0.23 & 3.81 & 3.81 & 17.0 & L
\enddata
\label{table:1b}
\end{deluxetable*}

\section{Results}
\label{results}
\subsection{The Data}
In principle, it is possible to remove the Faraday rotation contribution from the Galactic foreground (GRM) directly, by subtracting this from the RM, in order to obtain a Residual Rotation Measure (RRM), using $\textrm{RRM} = \textrm{RM} - \textrm{GRM}$. The RRM should consist of the extragalactic contribution to the line-of-sight Faraday rotation for each source. However, in practise, with current restrictions the RRM cannot be reliably calculated as they contain systematics from the foreground calculation alongside an additional inherent error term \citep[see][]{2015A&A...575A.118O}. To quote \citet{2015A&A...575A.118O}, \emph{``extragalactic contributions are not very well constrained by the data''} and \emph{``subtracting a Galactic foreground from the data is therefore not a good way of estimating extragalactic contributions''}. This inhibits analysis of the RRMs, unless some additional technique can be invoked to account for the extra uncertainties (see Section~\ref{conclusions}). With the current RM grid, this makes analysis of the GRMs a more reliable approach (as discussed further in Section~\ref{faradaybayes}). For our analysis, we use the GRM data from \citet{2015A&A...575A.118O}.

\begin{figure*}
\centering
\includegraphics[clip=false, width=0.49\hsize]{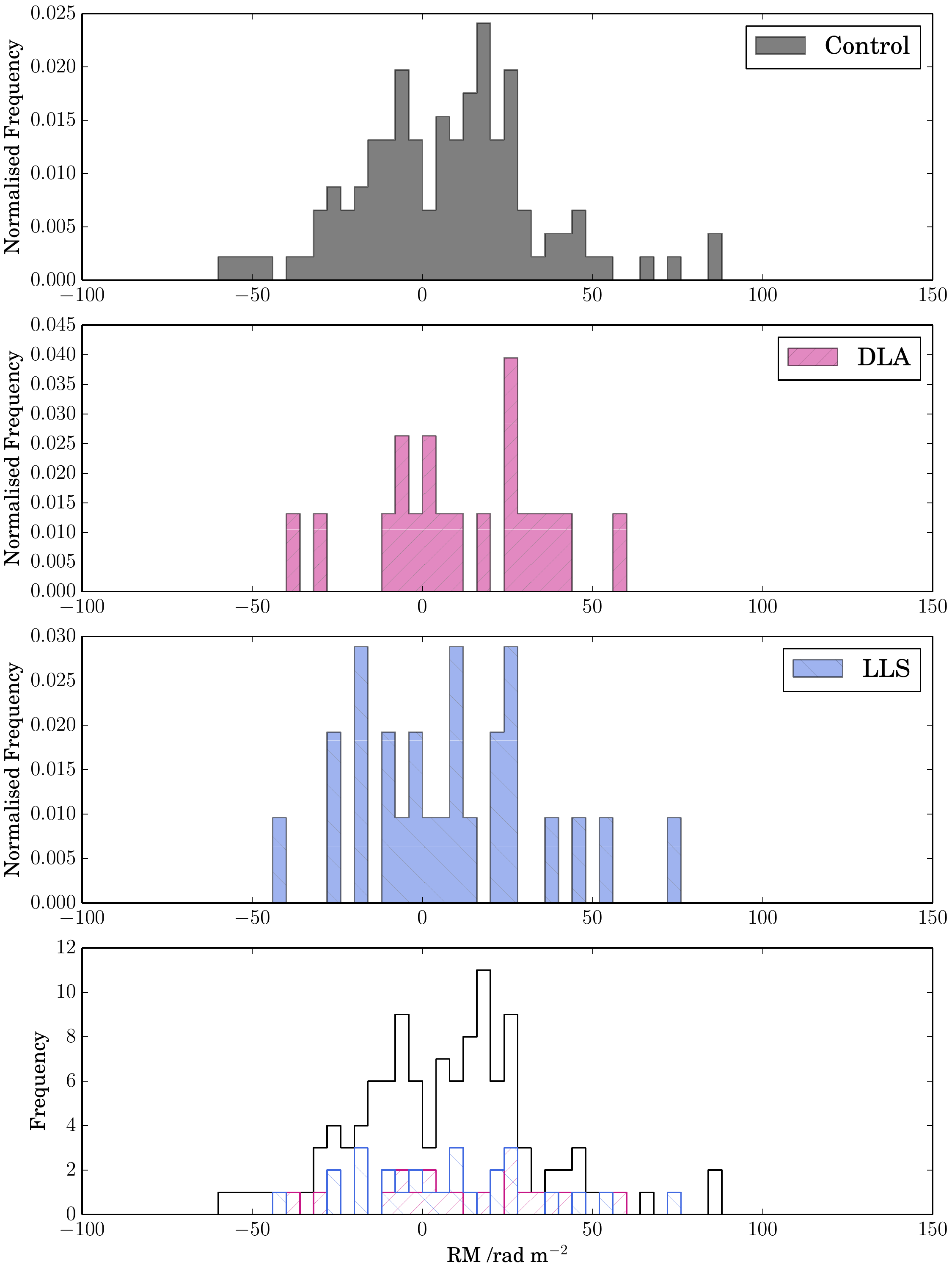}
\includegraphics[clip=false, width=0.49\hsize]{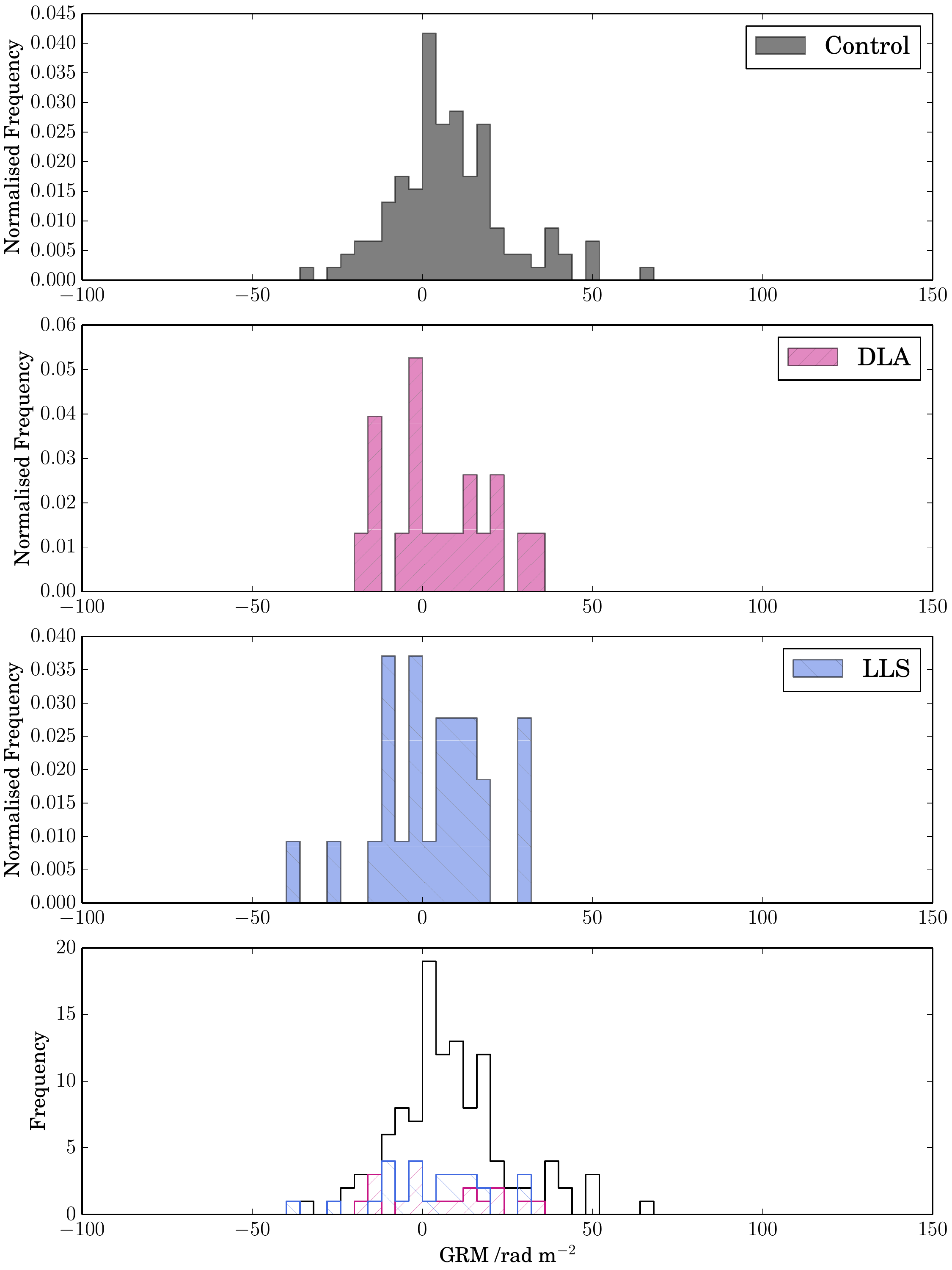}
\caption{\small{\textbf{Left column:} The RM distributions for sources with and without intervenors. The control sample (black, top row), the DLA sample (violet, 2nd row), and the LLS sample (blue, 3rd row) are all shown as normalised histograms. All three of the control, DLA, and LLS samples are also plotted simultaneously in non-normalised histograms, using the same colour-scheme (bottom row). 
\textbf{Right column:} The GRM distributions for sources with and without intervenors. The sample and colour-scheme is the same as in the left column.}}
\label{rmhistograms}
\end{figure*}

\begin{figure*}
\centering
\includegraphics[clip=false, width=0.498\hsize]{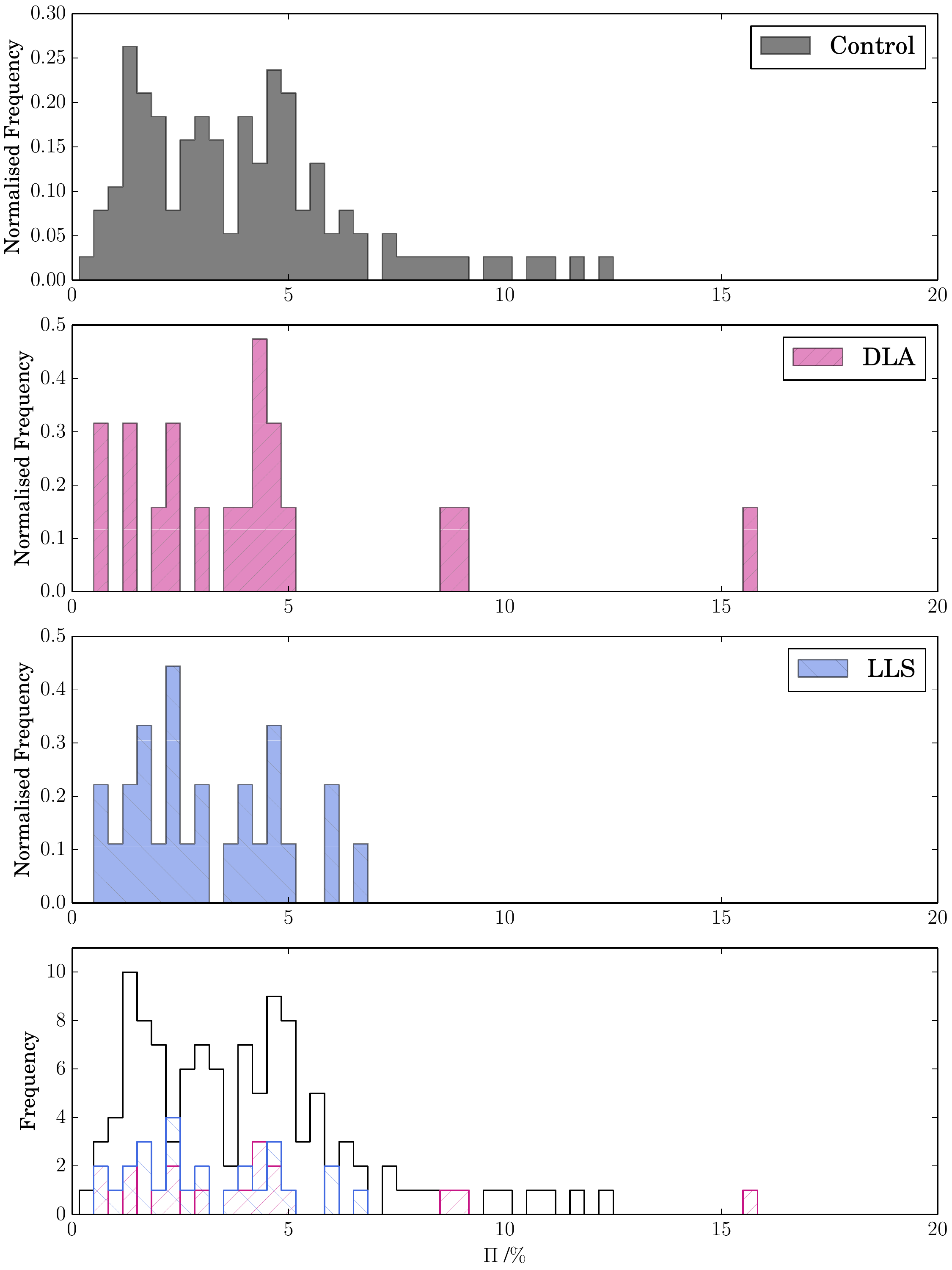}
\includegraphics[clip=false, width=0.498\hsize]{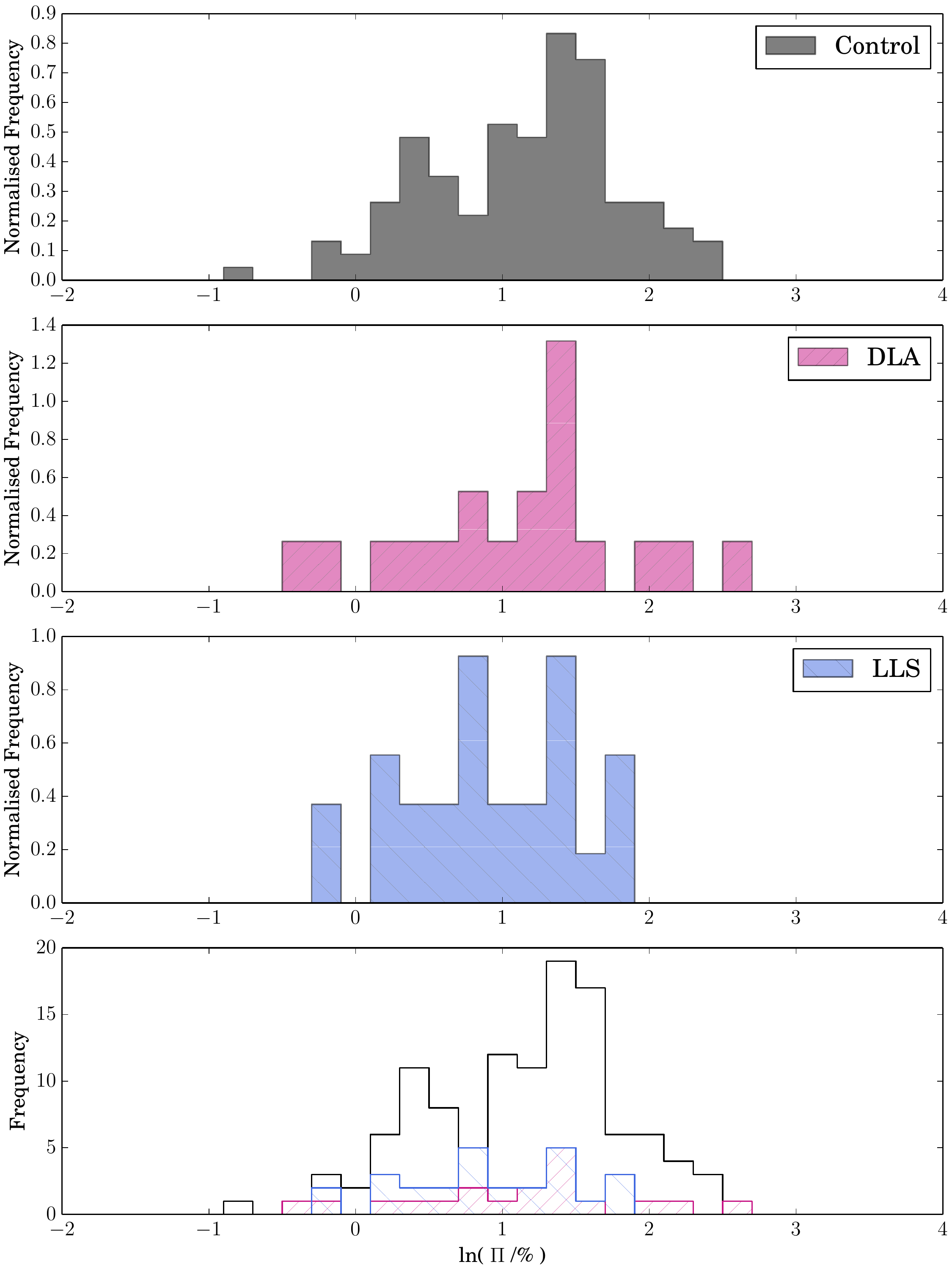}
\caption{\small{\textbf{Left column:} The polarized fraction, $\Pi$, distributions for sources with and without intervenors. The control (black, top row), the DLA (violet, 2nd row), and the LLS (blue, 3rd row) samples are all shown as normalised histograms. All three of the control, DLA, and LLS samples are also plotted simultaneously in non-normalised histograms, using the same colour-scheme (bottom row). \textbf{Right column:} The logarithm of the polarized fraction distributions for sources with and without intervenors. The samples and colour-scheme are the same as in the left column.}}
\label{pihistograms}
\end{figure*}

Our RM and GRM data are presented in Fig.~\ref{rmhistograms}. Our polarized fraction, $\Pi = \sqrt{Q^2+U^2}/I$, data are presented in Fig.~\ref{pihistograms}. The Figures show all three of our individual samples: the control, DLA, and LLS. The distributions of the various samples cannot be assessed reliably by eye, due to the small number statistics. We now will assess statistical differences between each of the samples using a full Bayesian framework. A Bayesian analysis is extremely advantageous for statistical analysis with low sample sizes, such as those often encountered in astronomy, as the analysis of small data sets can otherwise lead to statistical power issues and often suffer from biased parameter values. Similarly, in the `Big Data' era, fully Bayesian tests can avoid finding significant correlations in cases where none exist. A full introduction to Bayesian analysis is well-beyond the scope of this paper, and we refer the reader to the classic text of \citet{gelman2004} for extensive further details on the conventions of the method and notation. Extensive details on the Bayesian framework that we use, including the selection of our priors, is provided in Appendix~\ref{appendix-bayes}.

The analysis of the Faraday rotation (RMs and GRMs) is given in Section~\ref{faradaybayes}, while the analysis of the polarized fractions is given in Section~\ref{polfracbayes}. For our parameter estimates of the Faraday rotation, we estimate the parameters $\mu$ and $\sigma$, which correspond to the calculated mean and standard deviation of the Faraday rotation of our source population (which we model using a normal distribution). For our parameter estimates of the polarized fractions, we instead estimate the parameters $\textrm{E}[\Pi]$ and $\textrm{SD}[\Pi]$, which correspond to the arithmetic mean and arithmetic standard deviation of the polarized fractions of our source population (which we model using a log-normal distribution). In both cases, we estimate these parameters from the posterior using the 0.5 quantile, i.e.\ the median, and define the specified uncertainties with the 68\% credible interval. For extensive further details, please refer to Appendix~\ref{appendix-bayes}. On multiple occasions, we will refer to the typical RM and polarized fraction of each sample. The typical RM (and GRM) is parameterised by $\sigma$, which informs us of the typical magnitude of RM for each sample. As we here allow a sign dependence in RM, this is equivalent to the |RM| used in previous studies. The typical polarized fraction is parameterised by $\textrm{E}[\Pi]$, which informs us of the typical magnitude of the polarized fraction for each sample.

\subsection{Faraday Rotation}
\label{faradaybayes}
\begin{figure*}
\centering
\includegraphics[trim=0cm 1.5cm 0cm 0.0cm, clip=true, width=0.45\hsize]{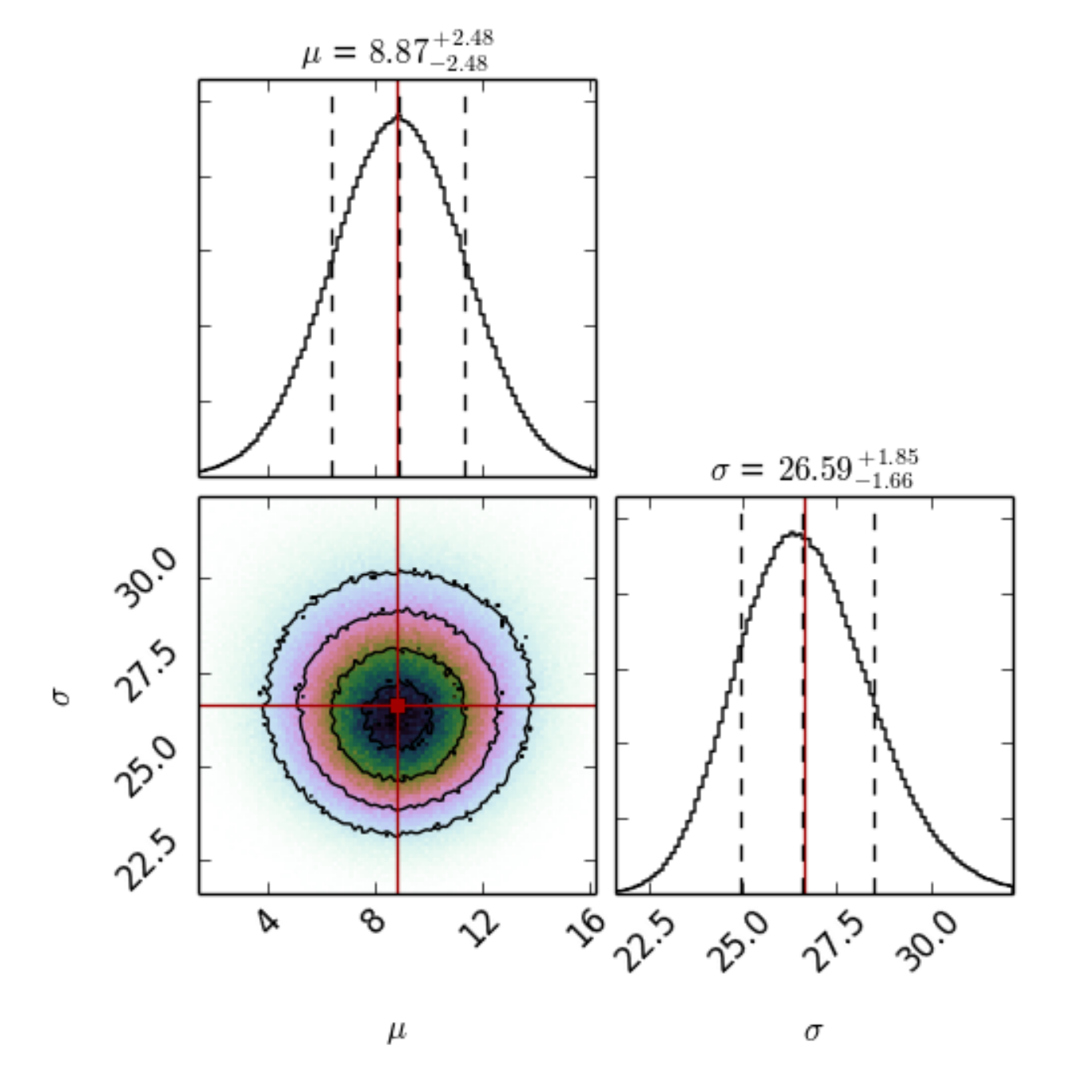}
\includegraphics[trim=0cm 1.5cm 0cm 0.0cm, clip=true, width=0.45\hsize]{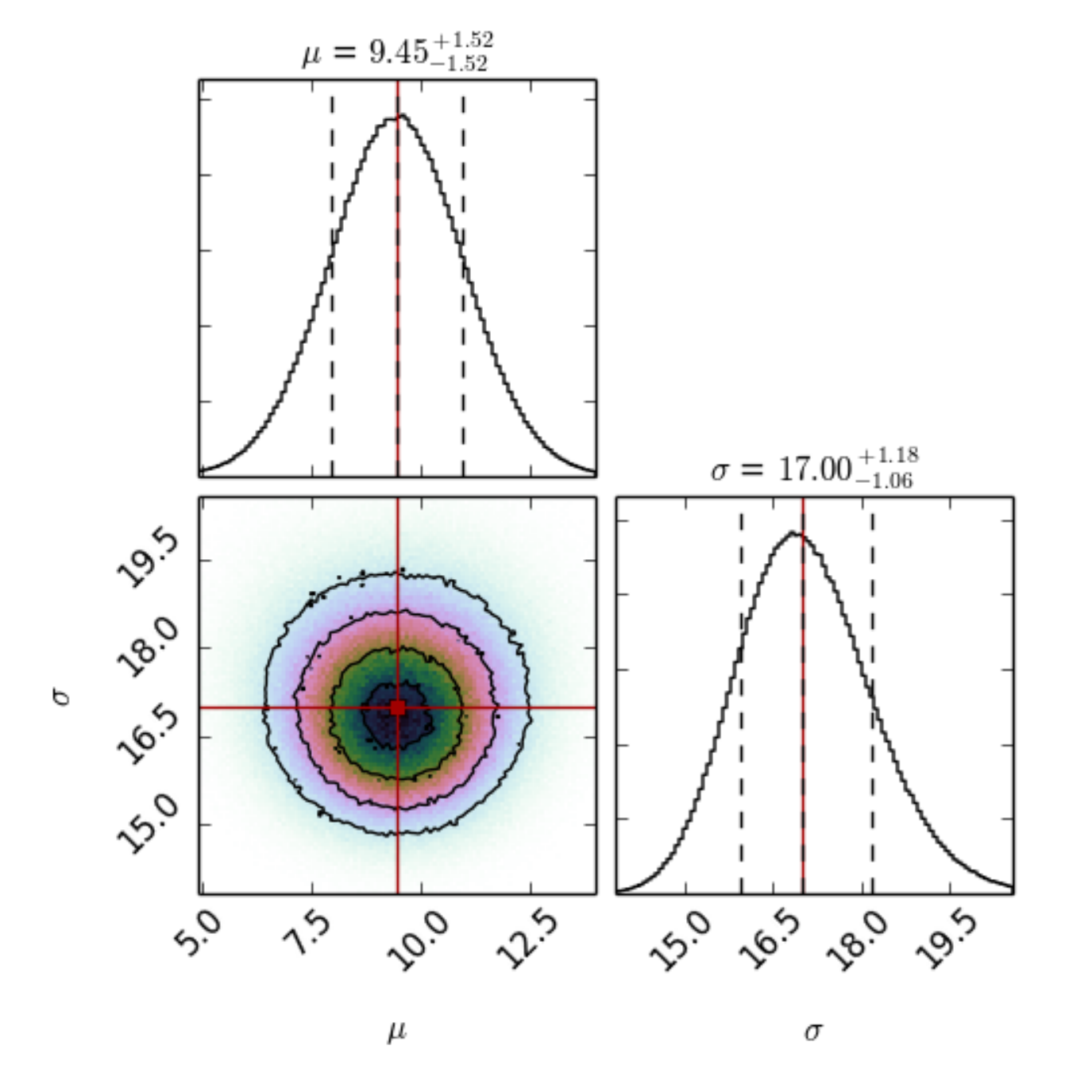} \\
\includegraphics[trim=0cm 1.5cm 0cm 0.0cm, clip=true, width=0.45\hsize]{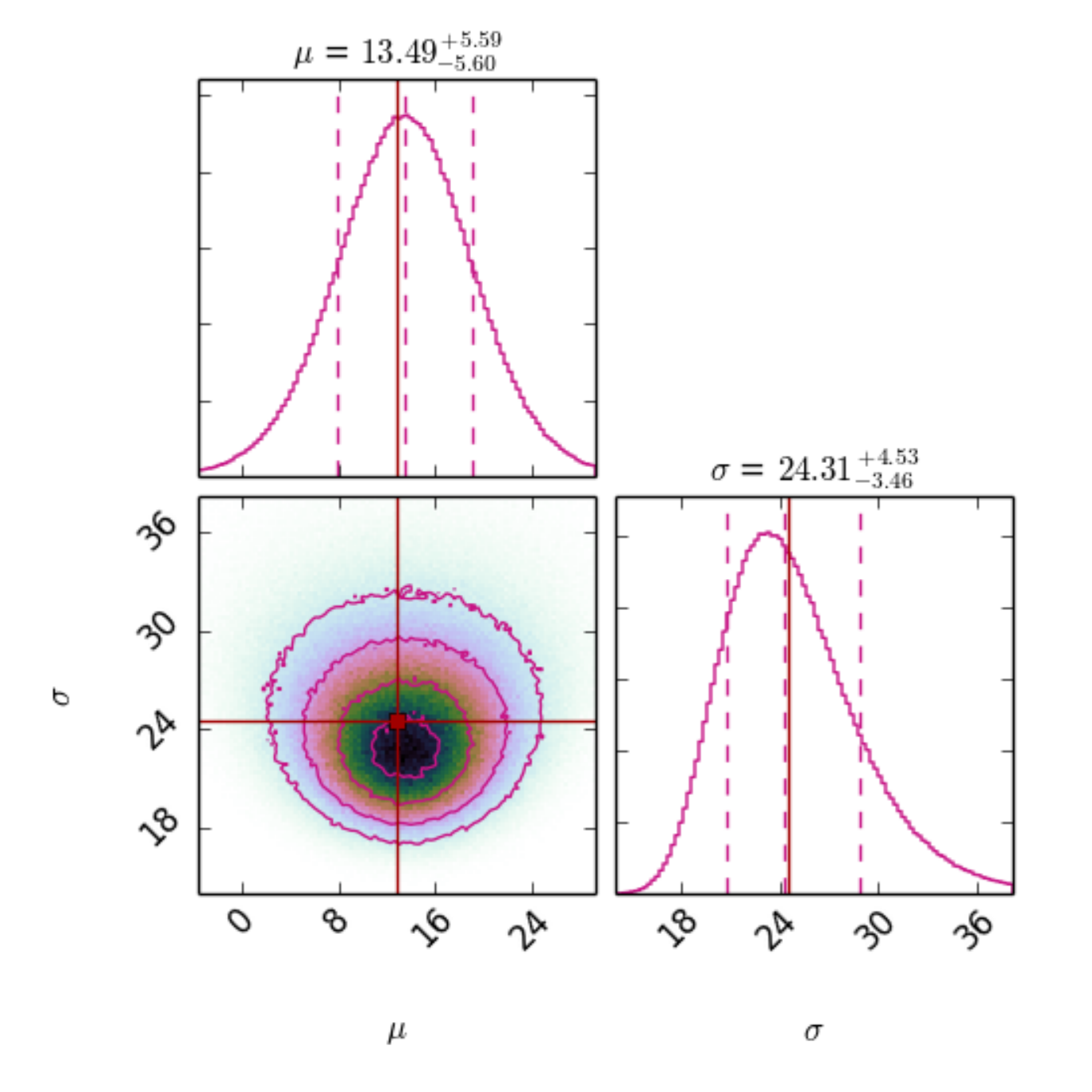}
\includegraphics[trim=0cm 1.5cm 0cm 0.0cm, clip=true, width=0.45\hsize]{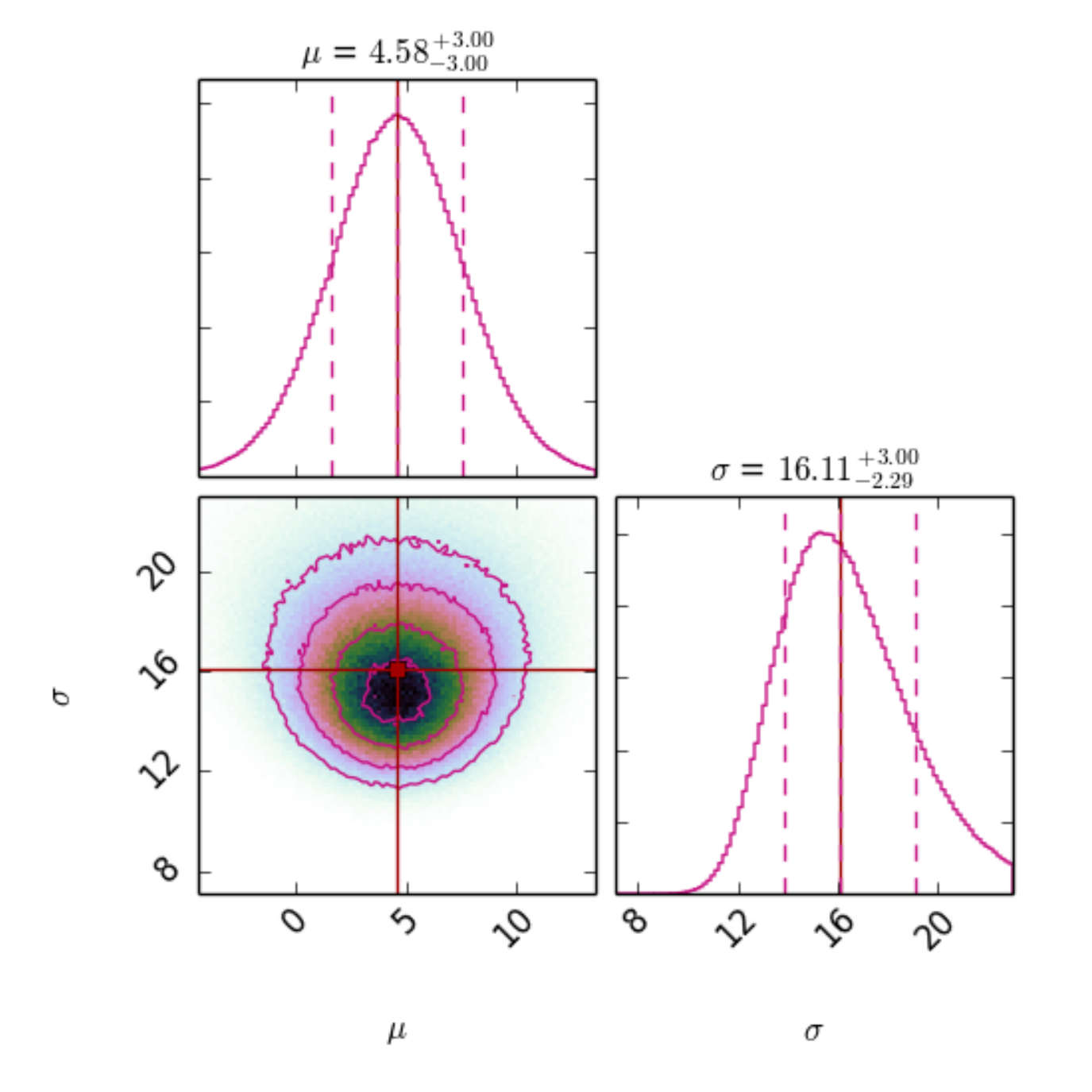} \\
\includegraphics[trim=0cm 0.0cm 0cm 0.0cm, clip=true, width=0.45\hsize]{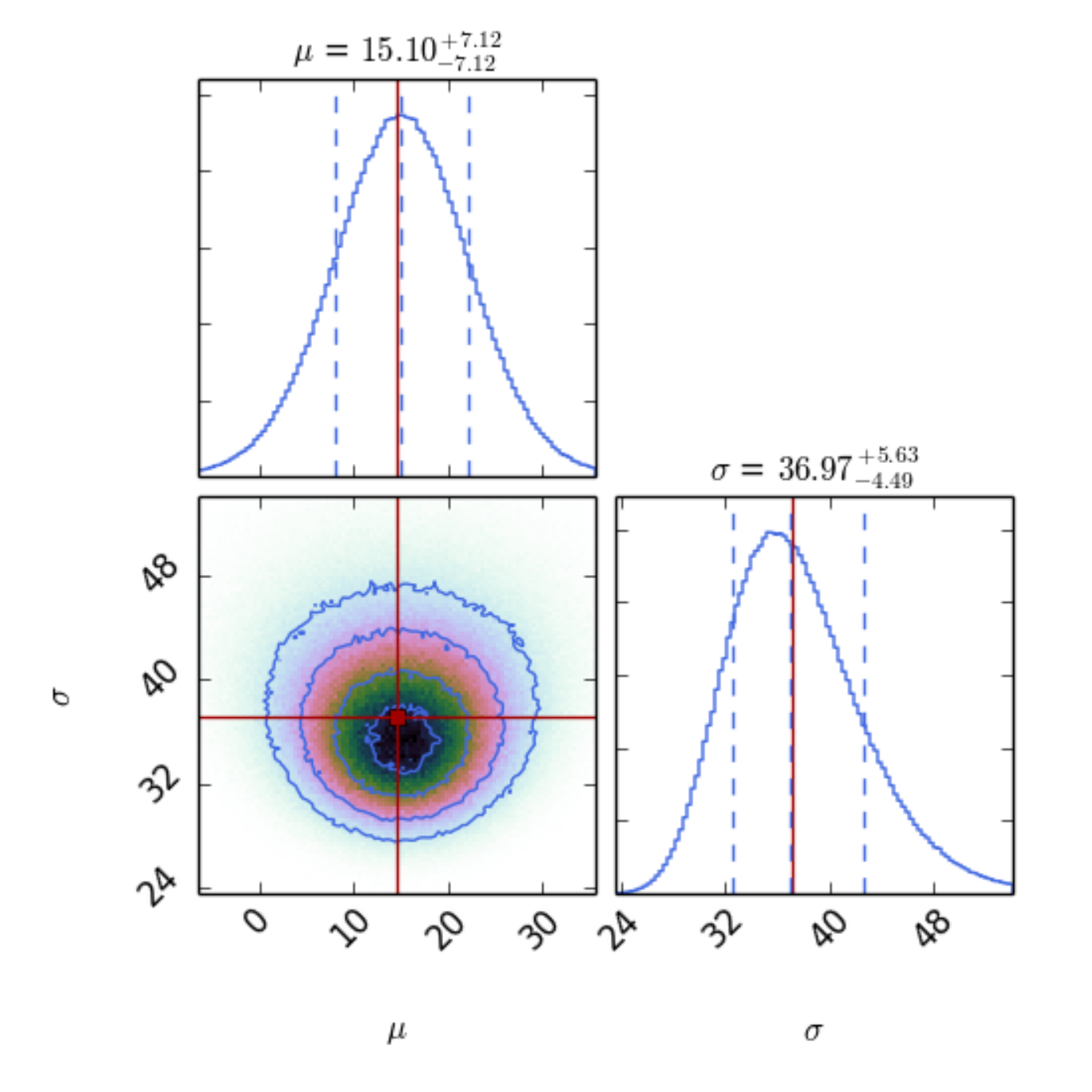}
\includegraphics[trim=0cm 0.0cm 0cm 0.0cm, clip=true, width=0.45\hsize]{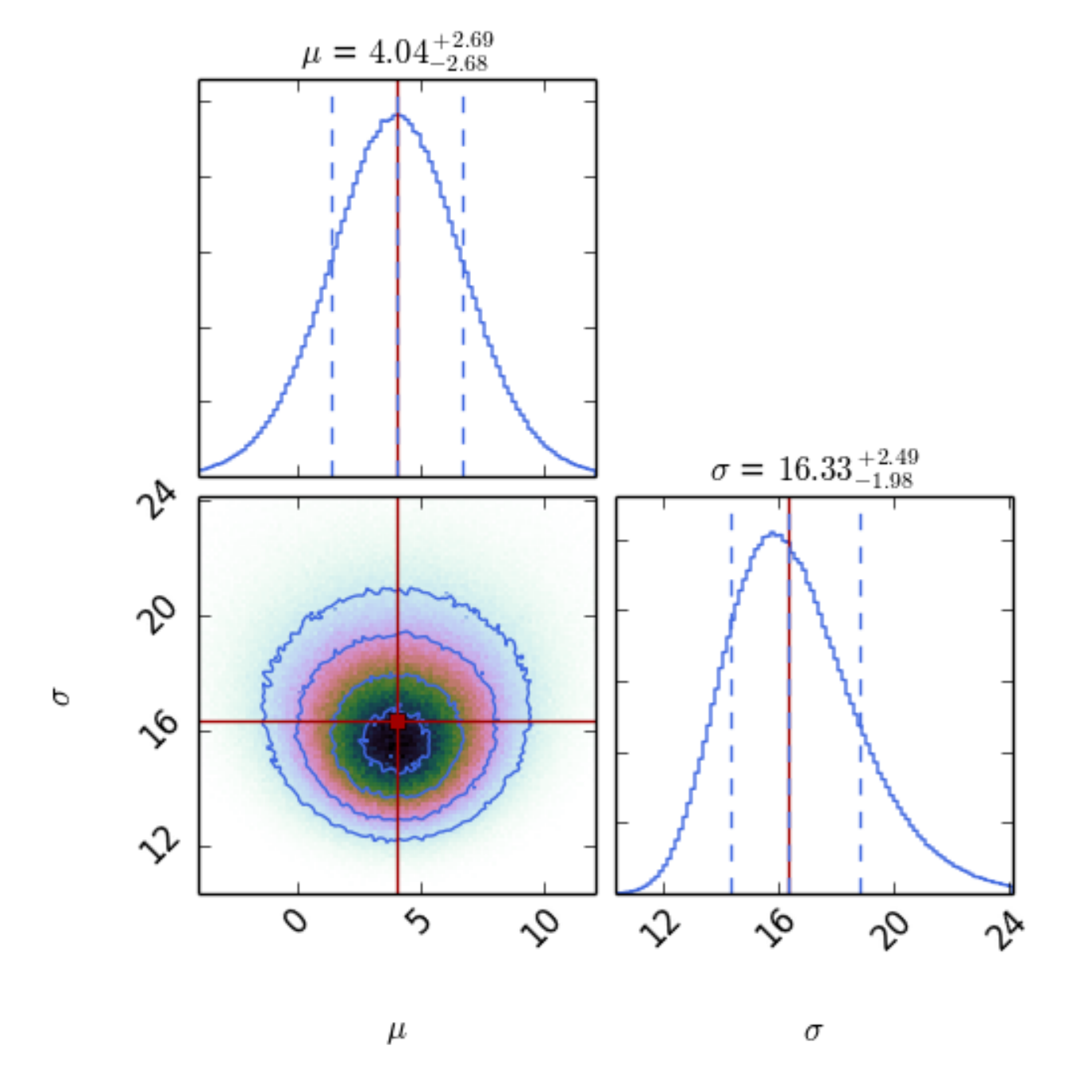}
\caption{\small{Parameter estimation of the RMs (left column) and the GRMs (right column). The estimates from the control (black, top row), the DLA (violet, middle row), and the LLS sample (blue, bottom row) are all shown. The parameter estimates for $\mu$ and $\sigma$ are stated in units of rad~m$^{-2}$. The estimated parameters are shown by the solid line, indicating the median, and the dashed lines, indicating the 68\% credible interval. The sample density is shown on a pseudocolour scale, and contour levels are shown at 11.7\%, 39.3\%, 67.5\%, and 86.5\%.}}
\label{estimationRMs}
\end{figure*}

We first consider the case of the RM and GRM values associated with the DLA and LLS. Without removal of the Galactic foreground, for a sample of sources that is evenly distributed spatially across the sky, we select a normal (Gaussian) distribution as our model for the RMs \citep[e.g.][]{2012arXiv1209.1438H,2014ApJS..212...15F}. However, a sample of sources that is not evenly distributed on the sky is expected to have a non-Gaussian distribution of RMs due to large-scale magnetic fields in the Milky Way that become imprinted onto the Faraday rotation signal of the background sources. It is therefore important to test that the Galactic foreground does not vary substantially between different samples. This Galactic foreground has been estimated in various studies \citep{2009ApJ...702.1230T,2012A&A...542A..93O,2015aska.confE..92J,2015A&A...575A.118O}, and can in principle be subtracted from the RM signal to yield an RRM (as discussed in Section~\ref{data}). Nevertheless, it has been suggested that while the Galactic foreground can be estimated using surveys such as the NVSS, it \emph{cannot} be reliably subtracted to obtain an RRM without knowing the relative uncertainties \citep{2015A&A...575A.118O}. We therefore check for Galactic variations using the GRMs, which are free from contamination by additional uncertainties. This methodology was also used in \citet{2014ApJ...795...63F} and is akin to stating that $\sigma_{\textrm{RM}}^2 = \sigma_{\textrm{GRM}}^2 + \sigma_{\textrm{RRM}}^2$. The sampled posterior distributions for the RMs and GRMs are shown in Fig.~\ref{estimationRMs}. 

To answer our scientific question, we want to know: what is the probability that the RMs of the control sample are larger than the RMs of the DLA (or LLS) sample, given our data? Or equivalently, for the DLAs,
\begin{equation}
p\left(\sigma[\textrm{DLA}]>\sigma[\textrm{control}]|D \right) ,
\label{theRMprob}
\end{equation}
where the probability is an evidential or Bayesian probability, rather than a frequentist probability.

We evaluate equation~\ref{theRMprob}, for the control, DLAs, and LLSs, through numerical integration by counting the number of posterior samples that satisfy e.g.\ $\sigma[\textrm{DLA}]>\sigma[\textrm{control}]$, and dividing by the total number of samples (2,000,000) drawn. The obtained probabilities are given in Table~\ref{table:2}.

\begin{deluxetable}{l c}
\tablewidth{0pt}
\tablecaption{Calculated $p(\textrm{RM}_{A}>\textrm{RM}_{B})$ and $p(\textrm{GRM}_{A}>\textrm{GRM}_{B})$ for the control, DLA, and LLS samples}
\tablehead{
\colhead{Probability} & \colhead{\%}     \\
\colhead{} & \colhead{}    }
\startdata
$\sigma_{\textrm{RM}}(\textrm{DLA})>\sigma_{\textrm{RM}}(\textrm{control})$     & 32.1         \\
$\sigma_{\textrm{RM}}(\textrm{LLS})>\sigma_{\textrm{RM}}(\textrm{control})$     & 99.0         \\
$\sigma_{\textrm{RM}}(\textrm{LLS})>\sigma_{\textrm{RM}}(\textrm{DLA})$         & 97.1       \\
$\sigma_{\textrm{GRM}}(\textrm{DLA})>\sigma_{\textrm{GRM}}(\textrm{control})$     & 38.2         \\
$\sigma_{\textrm{GRM}}(\textrm{LLS})>\sigma_{\textrm{GRM}}(\textrm{control})$     & 39.8         \\
$\sigma_{\textrm{GRM}}(\textrm{LLS})>\sigma_{\textrm{GRM}}(\textrm{DLA})$         & 52.6         
\enddata
\label{table:2}
\end{deluxetable}

\subsection{Polarized Fractions}
\label{polfracbayes}
\begin{figure}
\centering
\includegraphics[trim=0cm 1.8cm 0cm 0.0cm, clip=true, width=0.92\hsize]{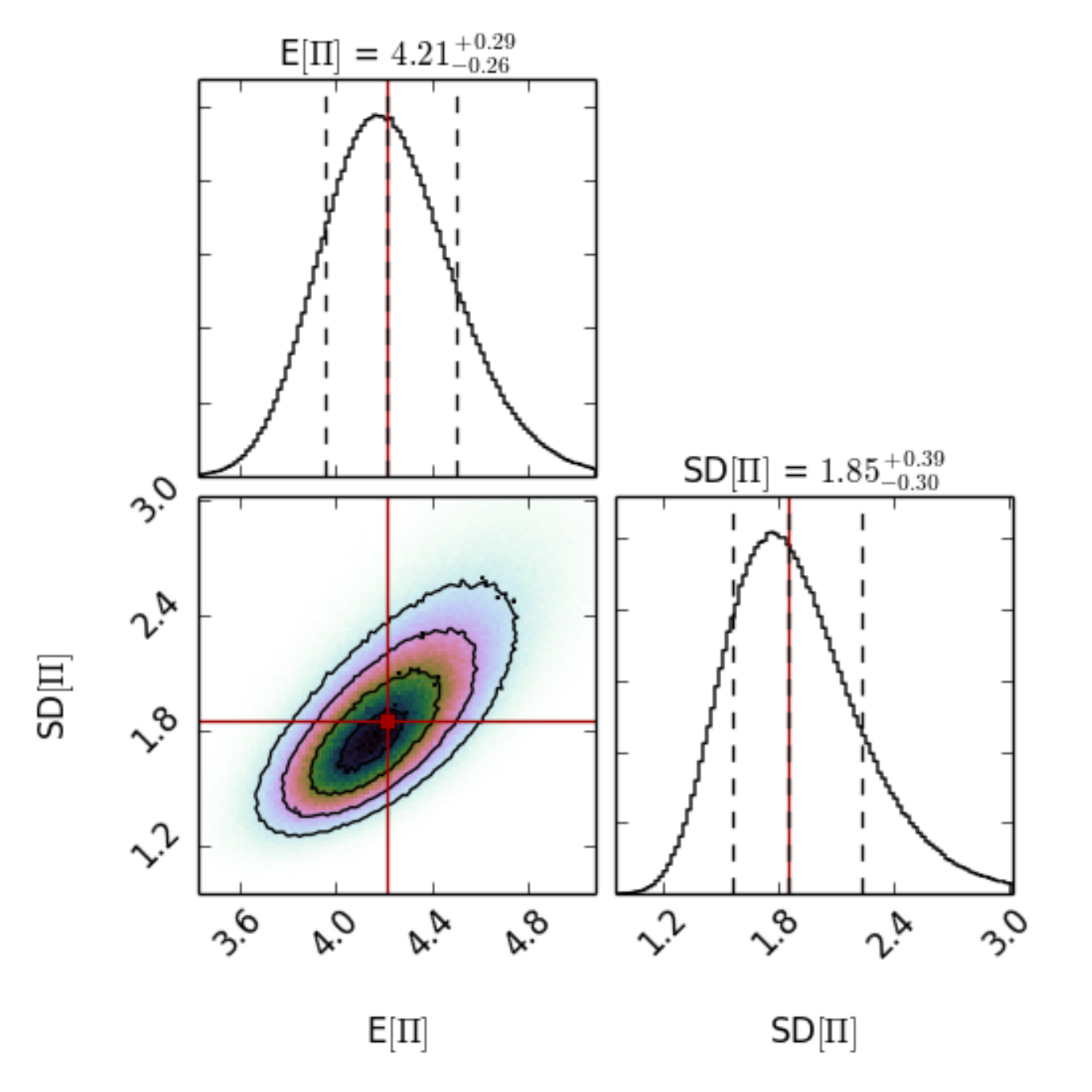}\\
\includegraphics[trim=0cm 1.8cm 0cm 0.0cm, clip=true, width=0.92\hsize]{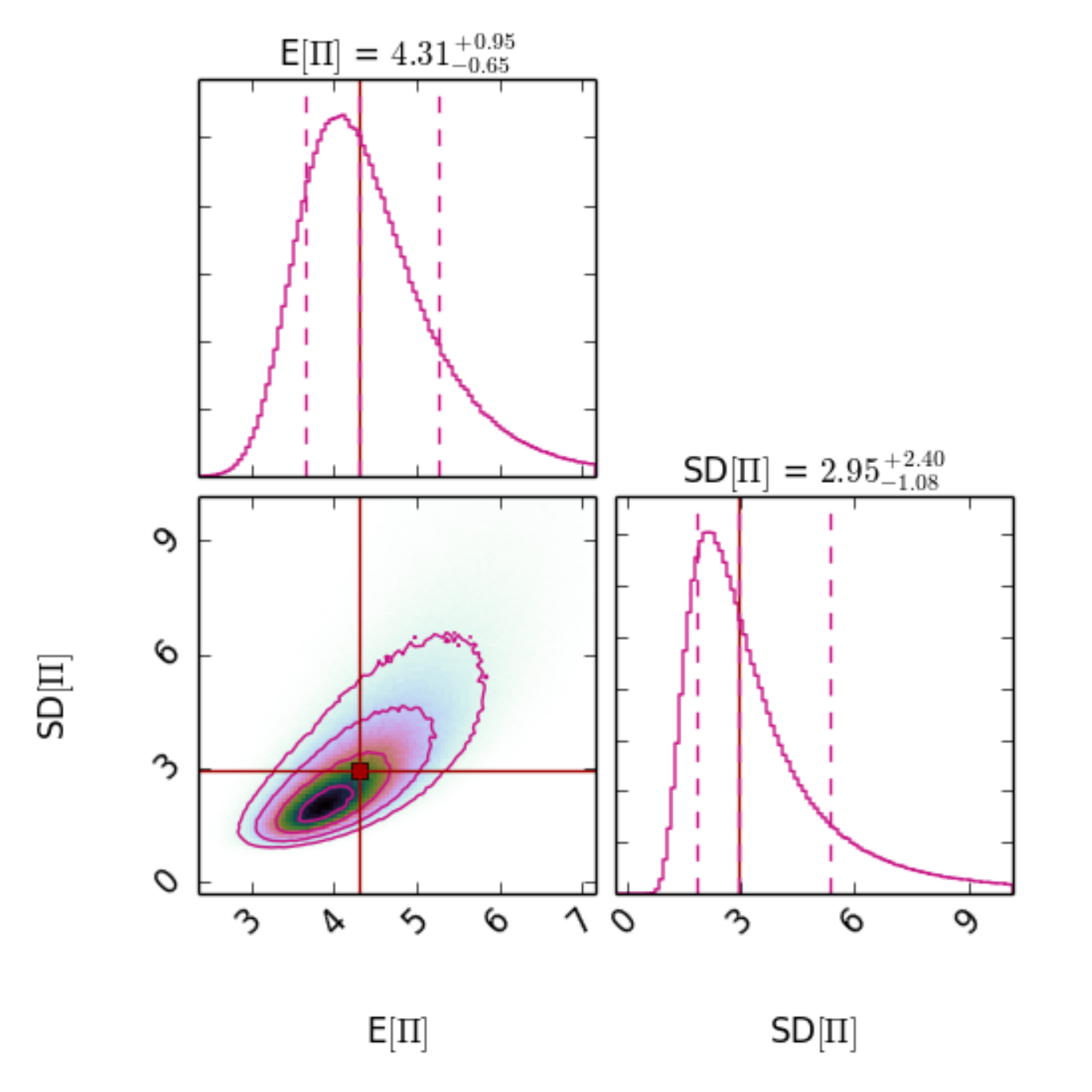} \\
\includegraphics[trim=0cm 0.0cm 0cm 0.0cm, clip=true, width=0.92\hsize]{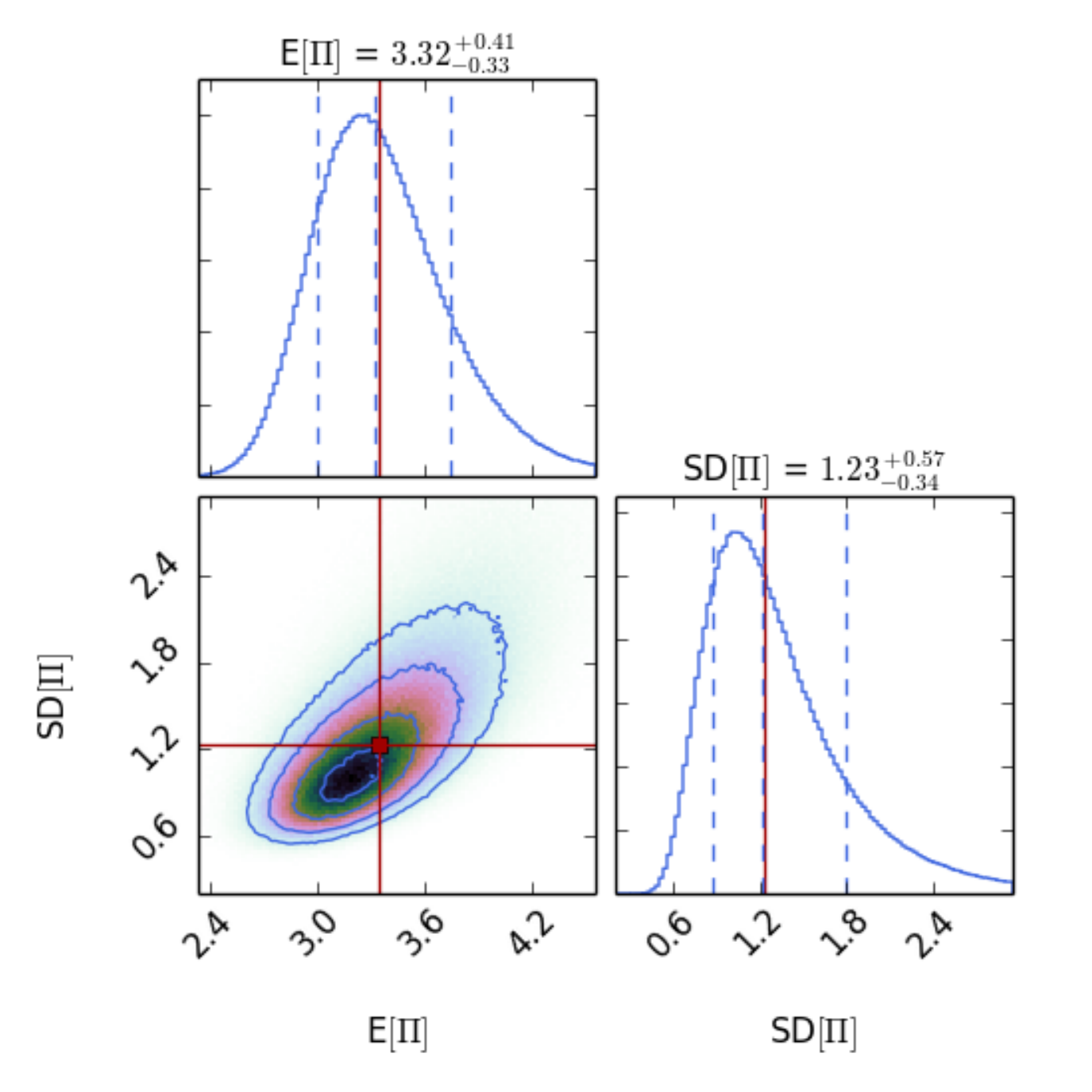}
\caption{\small{Parameter estimation of the polarized fractions, $\Pi$. The estimates from the control (black, top row), the DLA (violet, middle row), and the LLS sample (blue, bottom row) are all shown. The parameter estimates for $\textrm{E}\left[\Pi\right]$ and $\textrm{SD}\left[\Pi\right]$ are stated in units of \%. The estimated parameters are shown by the solid line, indicating the median, and the dashed lines, indicating the 68\% credible interval. The sample density is shown on a pseudocolour scale, and contour levels are shown at 11.7\%, 39.3\%, 67.5\%, and 86.5\%.}}
\label{estimationPIs}
\end{figure}

The sampled posterior distributions for the polarized fractions, $\Pi$, are shown in Fig.~\ref{estimationPIs}. 

To answer our scientific question, we want to know: what is the probability that the $\Pi$s of the DLA (or LLS) sample are smaller than the $\Pi$s of the control sample, given our data? Or equivalently, for the DLAs,
\begin{equation}
p\left(\textrm{E}[\Pi(\textrm{DLA})]<\textrm{E}[\Pi(\textrm{control})]|D \right) .
\label{thePIprob}
\end{equation}

We evaluate equation~\ref{thePIprob}, for the control, DLAs, and LLSs, through numerical integration, as described in Section~\ref{faradaybayes}. The obtained probabilities are given in Table~\ref{table:3}.

\begin{deluxetable}{l c}
\tablewidth{0pt}
\tablecaption{Calculated $p(\textrm{E}[\Pi_{A}]<\textrm{E}[\Pi_{B}])$ for the control, DLA, and LLS samples}
\tablehead{
\colhead{Probability} & \colhead{\%}     \\
\colhead{} & \colhead{}    }
\startdata
$\textrm{E}[\Pi(\textrm{DLA})]<\textrm{E}[\Pi(\textrm{control})]$    & 45.2        \\
$\textrm{E}[\Pi(\textrm{LLS})]<\textrm{E}[\Pi(\textrm{control})]$     & 95.5         \\
$\textrm{E}[\Pi(\textrm{LLS})]<\textrm{E}[\Pi(\textrm{DLA})]$         & 90.0            
\enddata
\label{table:3}
\end{deluxetable}

\subsection{Robustness Checks}
\label{robustness}
\subsubsection{Prior Selection}
Important tests of the validity of our results are robustness checks. The nature of robustness checks are covered in \citet{gelman2004}. The checks themselves do not modify our results, unless the check highlights some factor that drastically changes our conclusions. Firstly, we wish to ensure that our results are not unduly sensitive to the choice of prior, which is explained in detail in Appendix~\ref{appendix-bayes}. We expect no significant change, particularly as we have only used weakly-informative priors throughout our analysis. We therefore repeat our analysis with an uninformative prior, this time critically setting $\kappa_{0} = 0.0$ (equivalent to no prior measurements) to obtain a uniform distribution. Our results remain essentially unchanged, and the re-determined probabilities are given in Table~\ref{table:4}. Using the uninformative prior, we find a 30.8\% probability that the DLAs have lower RMs than the control, or equivalently a 69.2\% probability that the control have higher RMs than the DLAs. Note that this does not in any way modify our results, as this is a robustness check, and shows a change of only 1.3\%. In addition, the DLAs having lower RMs would be unphysical. This probability did not hold when we incorporated some physical information into the weakly-informative prior that we used for our actual analysis.

\subsubsection{Non-Intervenors}
\label{sensitivity}
As we have a sample of limited size, we cannot substantially subdivide or split the sample without losing meaningful statistical power. This limits our ability to explore other connections with these data, such as the evolution of magnetic fields with column density, redshift, or spectral index. However, to ensure that our results are correct, we can perform robustness checks on our data in which we exclude a small fraction of sources at a time to ensure a similar distribution of source parameters between all samples, and hence attempt to control for, and minimise any selection effects.

While measuring the magnetic environment of DLAs and LLSs is of interest, the primary motivation is in observing the \emph{intervening} systems that are unrelated to the background quasar itself. One robustness check requires ensuring that our results remain unchanged when only selecting the intervening DLA and LLS lines-of-sight. We therefore repeat our analysis, this time excluding the few sources with $\Delta z = z_{\textrm{qso}} - z_{\textrm{abs}} \le 0.04$. Our results remain essentially unchanged, and the re-determined probabilities are given in Table~\ref{table:4}.

\subsubsection{Spectral Distribution}
Following an equivalent rational to that in Section~\ref{sensitivity}, we also wish to ensure that our samples (the control, DLA, and LLS samples respectively) all have a similar distribution of spectral index. From the distribution of spectral index shown in Fig.~\ref{spectra}, it could be argued that the control sample show a slight enhancement of steep-spectrum sources with $\alpha\le-1.0$ in contrast to the DLA and LLS samples.

We therefore repeat our analysis, this time excluding the few sources with $\alpha\le-1.0$. Our results remain essentially unchanged, and the re-determined probabilities are given in Table~\ref{table:4}.

\subsubsection{Source Size}
Following an equivalent rational to that in Section~\ref{sensitivity}, we also wish to ensure that our samples (the control, DLA, and LLS samples respectively) all have a similar distribution of source size. From the distribution of effective area shown in Fig.~\ref{areas}, it could be argued that the control sample show a slight enhancement of more extended sources with $A\ge350$~arcsec$^2$ in contrast to the DLA and LLS samples.

We therefore repeat our analysis, this time excluding the few sources with $A\ge350$~arcsec$^2$. Our results remain essentially unchanged, and the re-determined probabilities are given in Table~\ref{table:4}.

\subsubsection{Redshift Distribution}
\label{redshiftdist}
Following an equivalent rational to that in Section~\ref{sensitivity}, we also wish to ensure that our samples (the control, DLA, and LLS samples respectively) all have a similar distribution of redshifts of the background QSOs, otherwise it could be argued that any difference between samples originates from differences in the selected QSOs. From the distribution of redshifts shown in Fig.~\ref{redshifts}, it could be argued that the DLA and LLS samples show an enhancement of sources with $z_{\textrm{qso}}\ge3.5$ in contrast to the control sample.

We therefore repeat our analysis, this time excluding the few sources with $z_{\textrm{qso}}\ge3.5$. The re-determined probabilities are given in Table~\ref{table:4}. The majority of our results remain essentially unchanged, however with two notable exceptions. Firstly, the probability that the RM of the DLAs is higher than the control hsa reduced to 17.4\%, or equivalently a 82.6\% probability that the control have higher RMs than the DLAs. Note that this does not in any way modify our results, as the DLAs having lower RMs would be unphysical. Secondly, the probability that the RM of the LLSs is greater than the control has reduced to $71.5$\%. This could be mild evidence for an enhanced RM associated with the LLSs, however it is a significant drop in probability of almost 30\% to when we do not control for the redshift distributions. We interpret this to indicate only a mild detection of enhanced RMs associated with the LLSs, and that we are most likely currently sensitive to the high-redshift distribution of QSOs. We therefore consider $71.5$\% to be the more conservative estimate.

\subsubsection{Sky Distribution}
The sources are distributed across the sky. As the magnetised Galactic foreground varies as a function of sky position, this makes the results sensitive to the sky distribution of sources. One possibility to explain the apparent difference between RMs of LLSs and a control sample, is that it is the consequence of contributions to the Faraday rotation from the Galactic foreground. Similarly, one could argue that the RM distribution is also responsible for the different polarized fractions. If the Galactic Faraday screen responsible for higher RM is also responsible for increased depolarization, then one would expect a lower polarized fraction for the LLSs. 

Currently the Galactic foreground cannot be reliably subtracted to obtain an RRM (see Section~\ref{faradaybayes}). Regardless, we expect the effect of the Galactic foreground to be low, as we have removed sources at low Galactic latitudes from our sample (see Section~\ref{data}) and we would expect sources with high RMs to be preferentially located in the Galactic plane \citep{2009ApJ...702.1230T}. For the foreground to be influencing our main results, our sample would have to be anisotropically distributed on the sky such that there was either a difference in the RM of sources with and without absorbing systems. Therefore if the Galactic foreground was causing our result, we could expect a different estimation of GRM between these different samples. 

We have found a 38.2\% probability that the GRM is different in the control and DLA samples. Similarly, we have found a 39.8\% probability that the GRM is different in the control and LLS samples. There is therefore no evidence that a varying Galactic foreground is influencing our results. The statistical differences in the RMs, with no associated difference in the GRMs, demonstrates that the LLSs cannot have significantly different sky distributions (relative to the Faraday structure of the Milky Way). We have therefore ruled out our results being affected by the sky distribution of sources.

\begin{deluxetable*}{l c c c c c c}
\tablewidth{0pt}
\tablecaption{Results of the robustness checks on calculated $p(\textrm{RM}_{A}>\textrm{RM}_{B})$, $p(\textrm{GRM}_{A}>\textrm{GRM}_{B})$, and $p(\textrm{E}[\Pi_{A}]<\textrm{E}[\Pi_{B}])$ for the control, DLA, and LLS samples. For clarity, our main results from Tables~\ref{table:2} and~\ref{table:3} are reproduced in the `Main Results' column.}
\tablehead{
\colhead{Probability} &  \colhead{Main Results}  & \colhead{Prior Selection}  & \colhead{Non-Intervenors}    & \colhead{Spectral Dist.}  & \colhead{Size Dist.}  & \colhead{$z$ Dist.}  \\
\colhead{} & \colhead{\%}  & \colhead{\%}  & \colhead{\%}  & \colhead{\%} & \colhead{\%} & \colhead{\%}  }
\startdata
$\sigma_{\textrm{RM}}(\textrm{DLA})>\sigma_{\textrm{RM}}(\textrm{control})$     & 32.1 & 30.8   & 43.3   & 43.0 & 35.8  & 17.4 \\
$\sigma_{\textrm{RM}}(\textrm{LLS})>\sigma_{\textrm{RM}}(\textrm{control})$     & 99.0 & 99.0   & 99.1   & 99.7  & 99.3  & 71.5 \\
$\sigma_{\textrm{RM}}(\textrm{LLS})>\sigma_{\textrm{RM}}(\textrm{DLA})$         & 97.1 & 97.2    & 95.0   & 97.0  & 97.1 & 87.8 \\
$\sigma_{\textrm{GRM}}(\textrm{DLA})>\sigma_{\textrm{GRM}}(\textrm{control})$     & 38.2 & 34.2  & 48.8   & 51.7   & 43.7  & 36.0 \\
$\sigma_{\textrm{GRM}}(\textrm{LLS})>\sigma_{\textrm{GRM}}(\textrm{control})$     & 39.8 & 39.5  & 46.0   & 55.6  & 46.3 & 46.0 \\
$\sigma_{\textrm{GRM}}(\textrm{LLS})>\sigma_{\textrm{GRM}}(\textrm{DLA})$         & 52.6 & 56.0   & 48.2  & 52.7  & 52.7   & 58.2 \\
$\textrm{E}[\Pi(\textrm{DLA})]<\textrm{E}[\Pi(\textrm{control})]$  & 45.2   & 36.5   & 40.5   & 42.0  & 44.6  & 34.7 \\
$\textrm{E}[\Pi(\textrm{LLS})]<\textrm{E}[\Pi(\textrm{control})]$   & 95.5  & 94.6   & 94.6   & 94.4   & 95.2 & 91.2 \\
$\textrm{E}[\Pi(\textrm{LLS})]<\textrm{E}[\Pi(\textrm{DLA})]$     & 90.0    & 91.6    & 90.3   & 90.1    & 90.1  & 90.5 
\enddata
\label{table:4}
\end{deluxetable*} 

\section{Discussion}
\label{discussion}
\subsection{The Nature of DLAs and LLSs}
\label{nature}
The fundamental nature of both DLAs and LLSs is uncertain. There have been several suggestions as to the physics and origin of these systems, often with varying degrees of certainty. Such studies have typically been limited by low sample sizes, uncertain statistical significance, and complicated interpretation of both statistical results (e.g.\ $p$-values) and of the physics (e.g.\ different galaxy types). However, while there is an emerging body of evidence that DLAs and LLSs definitively probe the distribution of collapsed, cold gas at high redshift and at different H\,{\sc i} column densities \citep[e.g.][]{1997ApJ...484...31G}, any related distinctions between the probed environments or galaxy types are still a matter of debate. Suggestions include that DLAs and LLSs may be related to different galaxy components such as disks versus halos, to high- versus low-impact parameters, or to clouds of turbulent halo gas versus more quiescent gas \citep{1993ASSL..188..263S,1995qal..conf.....M,2011MNRAS.416.1215R}. It has also been argued that photometry of higher column density DLAs lends support to the view that those systems are high-redshift galaxies \citep{1996Natur.382..234D,1996MNRAS.279L..27F}. Meanwhile, studies of lower column density systems similar to the LLSs indicates that those systems can even be associated with lines-of-sight that pass near galaxy groups or clusters \citep{1996ApJ...456L..17L}. It is of interest to investigate the connection between DLAs and LLSs, in order to better understand the links with galaxy formation. However, it is strongly likely that both systems provide insight into the nature of protogalaxies in the early Universe \citep{1999ApJ...511L..71P,2005ARA&A..43..861W}.

To date, our understanding of DLAs and LLSs has mostly been developed through observations, rather than through theory. This has arguably resulted in a relatively simple model of the physics that drives the evolution and growth of these objects, and resulted for example in possible abundance mismatches between simulations and observations \citep{2001ApJ...559..131G}. Nevertheless, the consistent consensus is that DLAs and LLSs are overdense regions in the intergalactic medium (IGM), and are therefore protogalactic clumps of some form \citep[e.g.][]{1999ApJ...511L..71P,2005ARA&A..43..861W}. Such systems likely constitute the predecessors of modern galaxies and have been proposed to be the progenitors of massive elliptical galaxies \citep{2003Natur.423...57P} and normal disk galaxies \citep{1996Natur.382..234D}. As such systems constitute significant neutral gas reservoirs and overdensities in the Universe, we therefore know that these systems must be connected to general galactic evolution. 

There is considerable evidence that DLAs are related to protogalaxies \citep{2000ApJ...545..603W}. More contemporary work \citep[e.g.][]{2015ApJ...808...38R} has also indicated that some fraction of DLAs occur in cold and dense inflowing streams (that extend over 10s of kpc) and that feed star formation in a massive central galaxy \citep{2008ApJ...683..149R,2011MNRAS.418.1796F,2012ApJ...748..121C}. There is also some evidence that DLAs can trace wind material that has been lifted away from galactic disks by star-formation driven outflows \citep{2008MNRAS.390.1349P,2008ApJ...683..149R}.

There is also a body of evidence that suggests LLSs probe at least three different types of system. The LLSs are known to be associated with similar systems to the DLAs \citep{1999ApJ...511L..71P}, and with the halos of low-mass galaxies \citep{2015MNRAS.451..904E}. They are also known to trace infall onto galaxies, being consistent with cold mode accretion streams \citep{2011ApJ...743..207R,2016MNRAS.455.4100F}. The same mechanism is also believed to be responsible for some DLAs \citep{2011MNRAS.418.1796F}. The LLSs have also been suggested to trace outflows, including those driven by star-formation and AGN activity, but also potentially including Galactic superwinds \citep{2006ApJ...648L..97P,2011Sci...334..952T}. As LLSs are clearly seen to trace phenomena that are associated with the circulation of matter into and out of galaxies, and as LLSs form a `bridge' between the higher column density DLAs and the lower column density Lyman-alpha forest, it is possible that LLSs may also trace the interface between protogalaxies and the intergalactic medium (IGM). This is supported by numerical simulations which find that LLSs occur on lines-of-sight that pass through the outer parts of more massive protogalaxies (and also near the centre of younger, lower density systems) \citep{1996ApJ...457L..57K}. Either way, the LLSs are potentially associated with the medium surrounding galaxies, e.g.\ the circumgalactic medium.

Consequently, DLAs and LLSs offer potential probes to trace the evolution of these various phenomena throughout cosmic time. Furthermore, the study of protogalactic clumps, and consequent galactic building blocks, potentially opens a window for studying magnetic fields in the cosmic web. As both systems are known to have some relation to infall and outflows, there is also an overlap with the more well-studied Mg\,{\sc ii} systems \citep[e.g.][]{1992ApJS...80....1S,1997ApJ...480..568S,2007ApJ...662..909K,2007ApJ...669..135M}. There is even some evidence to suggest that strong Mg\,{\sc ii} systems and LLSs probe similar physical systems \citep{1993ASSL..188..263S,1999ApJS..120...51C}, and consequently LLSs would extend the measurement of strong Mg\,{\sc ii} systems out to higher redshifts. Nevertheless, while this is difficult to reconcile with the very different column densities probed by such systems, it is likely that these various absorption line systems have considerable overlap and can probe a range of different environments. However, via QSO-absorption line studies, we can still study the \emph{average} system represented at each respective column density range. It is likely that a considerable leap in sample size would be required, in order to clearly identify the overlaps between the physical systems measured at different column densities (and also to avoid any arbitrary distinctions based on column density). This is especially true as each physical system also likely evolves with redshift and other parameters which ideally require isolation.

\subsection{Ionised Gas in DLAs and LLSs}
\label{ionisation}
We are interested in proxies for characteristics of the magnetic field -- in particular measurements of Faraday rotation, via the RM, and of magnetic field disorder or free electron variations, via the polarized fraction (which is itself a proxy for the depolarization). In particular, measuring an RM fundamentally requires the presence of ionised or partially ionised gas that is permeated by magnetic fields.

Understanding how this ionization affects our measurements, requires consideration of the overlap with the physical nature of DLAs and LLSs as detailed in Section~\ref{nature}. Are the DLAs and LLSs physically different from one another, or are we imposing arbitrary distinctions due to the column density criterion? There is a critical and fundamental distinction between both systems: the hydrogen gas is mostly neutral in DLAs, whilst the gas is ionised in other absorption systems such as LLSs.

The neutrality of the gas in DLAs is crucial to their important role as neutral gas reservoirs for star-formation in the high-redshift Universe \citep{2005ARA&A..43..861W}. The birthplace of stars -- molecular clouds -- are typically preceded and formed from cold \emph{neutral} clouds, from which stars are able to form. However, stars are unlikely to form from warm \emph{ionised} gas. The DLAs therefore enable a view onto the interaction between neutral gas and new star-formation, and are potentially the only example of an interstellar medium (ISM) at high redshifts. DLAs also remain predominantly neutral due to `self-shielding', which prevents ionisation of the gas. While ultraviolet radiation ionises the gas, at these extreme high column densities the ionised material rapidly recombines and becomes neutral again \citep{2000ApJ...545..603W}.

At lower column densities, it is unclear where the transition from predominantly neutral to predominantly ionised gas occurs, except that it is below the defining column density of DLAs (log $N$(H\,{\sc i})$\ge20.3$) \citep{1997ApJ...487...73P,1998ApJ...495..647H,2001MNRAS.326.1475M}. In both the LLSs (17 < log $N$(H\,{\sc i}) < 20.3), and the Ly-alpha forest (log $N$(H\,{\sc i})$\le17$), the neutral gas is a minor or non-existent phase, and so these lower column density systems primarily trace ionised gas. However, as the H\,{\sc i} column density increases from the threshold of a LLS up to a DLA, the systems go from mostly ionised to mostly neutral due to the self-shielding \citep{2015MNRAS.451..904E}. As such, the majority of the hydrogen gas in DLAs is typically considered to be completely neutral (with a negligible ionisation fraction) in the H\,{\sc i} phase \citep{2009MNRAS.397.2037M}. 

Nevertheless, there is actually a small amount of ionized gas in DLAs, as evidenced by the highly ionised metal absorption lines seen in many DLAs. For example, \citet{2008MNRAS.390....2L} found that the ionisation fraction H\,{\sc ii}/H is about 12 to 20\% in DLAs. Radio observations of ionised gas in the Galactic disk of the Milky Way have shown that the mass ratio of H\,{\sc ii} to H\,{\sc i} gases is $\approx0.01$ \citep{1989agna.book.....O,2005ApJ...623...99O}. Meanwhile, \citet{2002ApJ...571..693P} and \citet{2011A&A...534A..82F} reported a hydrogen ionisation fraction in DLAs of $\bar{x}$(H\,{\sc i}) = 10--50\%. Furthermore, \citet{1996ApJ...470..403P} calculated the ionisation fraction in DLAs by taking radiation transfer into account, and assuming an ultraviolet background intensity corresponding to that at $z\sim2$--3. They found the ionisation fraction $\bar{x}$($n_{e}/n$) = 4\%. 

In contrast, the LLSs definitively trace ionised gas. In one system, an ionisation fraction of $97\pm2$\% was reported \citep{1999ApJ...511L..71P}. Meanwhile, the LLSs with highest column densities (19 < log $N$(H\,{\sc i}) < 20.3), sometimes referred to as the Super Lyman Limit Systems (SLLS), have been found to have mean ionisation fractions of 90\% \citep{2007MNRAS.382..177P,2009ASSP...10..419P}. Furthermore, a study of 157 systems confirmed that the majority of LLSs are highly ionised, and also showed that there is an increasing ionisation state of the gas with decreasing column density \citep{2015ApJS..221....2P}.

\subsection{Coherent Magnetic Fields}
\label{coherent}
We can use our measurements of the RM and GRM associated with the absorption line systems, relative to the control sample, in order to place constraints on the coherent magnetic field strength in the DLAs and the LLSs. We assume throughout that any change in the data resembles a change in the overall population, rather than the emergence of a significant number of physical outliers in one of the samples. Such a scenario could occur if for example there were two populations of sources in one of our samples, perhaps one population having undergone significant dynamo activity and the other population having not. Such an alternative could only be addressed with much better statistics.

Following \citet{1982ApJ...263..518K}, for an RRM excess measured in the observing-frame, $\Delta \textrm{RRM}$, which arises in the rest-frame at redshift $z$,
\begin{equation}
\Delta \textrm{RRM} = \beta (1+z)^{-2} N_{e} \left< B_{\parallel} \right> \\, 
\end{equation}
where $N_{e}$ is the electron column density in cm$^{-2}$, $\beta = 2.63\times10^{-19}$~rad~m$^{-2}$~cm$^{-2}$~$\muup$G$^{-1}$, and $\left< B_{\parallel} \right>$ for any field reversal pattern is
\begin{equation}
\left< B_{\parallel} \right> = \frac{\int n_{e} B_{\parallel} dl}{\int n_{e} dl} \\,
\end{equation}
where $n_{e}$ is the number density of free electrons in cm$^{-3}$, $B_{\parallel}$ is the component of the magnetic field parallel to the line-of-sight in $\muup$G, and $dl$ is a finite element of the path length in pc. The integrals are defined from us to the background source. The line-of-sight component of an intervenor's magnetic field is therefore given by
\begin{equation}
\left< B_{\parallel} \right> = \frac{\Delta \textrm{RRM}(1+z)^{2}}{\beta N_{e}} \\.
\end{equation}

The electron column density is related to the neutral hydrogen column density, $N(\textrm{H\,{\sc i}})$, by
\begin{equation}
N_{e} \approx \frac{\bar{x}}{(1-\bar{x})} N(\textrm{H\,{\sc i}}) \\,
\end{equation}
where $\bar{x}$ is the hydrogen ionisation fraction. Consequently, we can write
\begin{equation}
\left< B_{\parallel} \right> = \frac{(1-\bar{x}) \Delta \textrm{RRM}(1+z)^{2}}{\beta \bar{x} N(\textrm{H\,{\sc i}})} \\.
\end{equation}

In our calculations, we use a small ionisation fraction of 5\% in DLAs, and an ionisation fraction of 95\% in LLSs. These estimates are justified in Section~\ref{ionisation}. For DLAs, the low ionisation state limits the expected RRM excess, however the very high column density ensures that even a small ionisation fraction can result in a large electron density along the line-of-sight. Note that we cannot measure the RRM excess directly (see Section~\ref{faradaybayes}), however, we have shown that there is no evidence for GRM differences between the samples, and can therefore use the RM excess as a proxy.

For the DLAs, given our data, there is a 32.1\% probability that the RM is larger in the DLAs than in the control. There is therefore little evidence for enhanced coherent magnetic fields in the DLAs. If there were an increase, we calculate that there is a 90\% probability that any increase over the control is, in RM, $<1.88$~rad~m$^{-2}$ in the observing-frame. For our sample, we have a median log[$N$(H\,{\sc i})]=20.63, and a median DLA redshift of 1.947. We therefore estimate, with 90\% probability (given our data), that the regular coherent magnetic fields within the DLAs must be $\le2.8$~$\muup$G. This estimate depends on the redshift range of the data, which varies between $z=0.53$--2.22. This corresponds to magnetic field upper limits of $\le0.8$--3.3~$\muup$G. The estimate further depends on the $N$(H\,{\sc i}) range of the data which varies between $2\times10^{20}$--$5\times10^{21}$ (by a factor of 25). Using the 25th and 75th percentiles, this would then correspond to magnetic field upper limits of $\le1.7$--11.8~$\muup$G.

For the LLSs, given our data and when controlling for the redshift distribution (see Section~\ref{redshiftdist}), there is a 71.5\% probability that the RM is larger in the LLSs than in the control. There is therefore mild evidence for enhanced coherent magnetic fields in the LLSs. We calculate that there is a 90\% probability that any increase over the control is, in RM, $<10.4$~rad~m$^{-2}$ in the observing-frame. For our sample, we have a median log[$N$(H\,{\sc i})]=18.58, and a median LLS redshift of 1.082. We therefore estimate, with 90\% probability (given our data), that the regular coherent magnetic fields within the LLSs must be $\le2.4$~$\muup$G. This estimate depends on the redshift range of the data, which varies between $z=0.62$--3.27. This corresponds to magnetic field upper limits of $\le1.4$--9.9~$\muup$G. The estimate further depends on the $N$(H\,{\sc i}) range of the data which varies between $1\times10^{17}$--$2\times10^{20}$ (by a factor of 1995). Using the 25th and 75th percentiles, this would then correspond to magnetic field upper limits of $\le0.8$--33~$\muup$G.

One could argue that an alternative averaging procedure would be more appropriate, whereby we calculate the magnetic field in each absorber individually, and instead average the values of $\left< B_{\parallel} \right>$. This would better incorporate the very large ranges in $N(\textrm{H\,{\sc i}})$ and the variation arising from the factor $(1+z)^{2}$. Re-averaging provides updated mean estimates for coherent fields in DLAs to $\le2.8$~$\muup$G (unmodified from the original estimate) and in LLSs to $\le47.2$~$\muup$G, and due to significant scatter provides median estimates in DLAs of $\le2.1$~$\muup$G and in LLSs of $\le1.8$~$\muup$G. However this alternative averaging does not include variations in $\bar{x}$ or, crucially, $\textrm{RRM}$ between sources. As we have measured an ensemble average of $\textrm{RRM}$, we believe it more correct to provide an ensemble average of $\left< B_{\parallel} \right>$, and therefore rely on our former estimates.

These prior upper limits are magnetic field constraints on the ISM of the DLAs/LLSs themselves. In the case of the DLAs, there may also be ionised material surrounding the high-density (mostly neutral) material, in a more highly-ionised halo gas component. For example, C\,{\sc iv} properties imply a ubiquitous, highly ionised, and enriched medium that traces the environments surrounding DLAs \citep{2015ApJ...808...38R}. These C\,{\sc iv} halos could give rise to Faraday rotation associated with the DLAs, even if the high-column density component of the DLA itself is free of ionisation. Following \citet{2015ApJ...808...38R}, the high incidence of strong C\,{\sc iv} absorption and the large scales over which it is distributed point to a substantial reservoir of metals in the diffuse material surrounding DLAs at $z\sim2$. A typical column density for C\,{\sc iv} in DLAs is $N(\textrm{C\,{\sc iv}}) = 10^{14}$~cm$^{-2}$ \citep{2007A&A...473..791F}, and a conservative ionisation fraction is $\bar{x}(\textrm{C\,{\sc iv}}) = 0.3$ \citep{2007A&A...465..171F}, which has been demonstrated to be the maximum possible ionisation fraction in models assuming either photo- or collisional ionisation. Taking there to be a 90\% probability that any increase in Faraday rotation of the DLAs over the control is, in RM, $<1.88$~rad~m$^{-2}$, and assuming that all of this Faraday rotation occurs in a C\,{\sc iv} halo, then this places a very weak and uninformative constraint on the coherent magnetic field strength of $\le1.5$~G. This is a magnetic field constraint for the typical halo surrounding a DLA.
 
Given no information on the ionisation state of the DLAs or the LLS, the Faraday rotation measurements can also serve as an indicator of the ionisation fraction of these systems. Making a reasonable assumption of a universal coherent magnetic field strength of $1$~$\muup$G within the absorber, then this is consistent with a 90\% probability that the DLA ionisation fraction is $\bar{x}\le12.7$\% and that the LLS ionisation fraction is $\bar{x}\le97.8$\%.

\subsection{Random Magnetic Fields \& Turbulence}
Foreground intervening screens that contain ionised gas and turbulent magnetic fields are known to depolarize a background source \citep{1966MNRAS.133...67B,1998MNRAS.299..189S,2009ApJ...693.1392S}. Such depolarization manifests as a reduction in the observed polarized fraction. We can therefore use our measurements of the polarized fractions associated with the absorption line systems, relative to the control sample, in order to place constraints on the random magnetic fields, the ionised gas distribution, and the degree of turbulence in the DLAs and the LLSs.

Following \citet{1966MNRAS.133...67B}, for the case of external Faraday dispersion by a non-emitting screen that contains thermal electrons and turbulent magnetic fields, but does not contain cosmic ray electrons, the depolarization is given by
\begin{equation}
\Pi = \Pi_{0} e^{-2\sigma_{\textrm{RM}}^2 \lambda^4} \\,
\end{equation}
where $\sigma_{\textrm{RM}}$ is the RM dispersion from the foreground screen(s), and $\Pi$ and $\Pi_{0}$ are the observed and intrinsic polarized fractions respectively, and $\lambda$ is the observing frequency. We wish to compare a sample, $k$, which contains lines-of-sight with both an intrinsic and an intervening screen, such that
\begin{equation}
\sigma_{\textrm{RM},k}^2 = \sigma_{\textrm{RM,intrinsic}}^2 + \sigma_{\textrm{RM,interv}}^2 \\,
\end{equation}
with a control, $j$, which contains lines-of-sight with only an intrinsic screen, such that
\begin{equation}
\sigma_{\textrm{RM},j}^2 = \sigma_{\textrm{RM,intrinsic}}^2 \\.
\end{equation}

Since $\Pi_{0}$ is the intrinsic polarized fraction of the background QSOs, it can therefore be trivially shown that
\begin{equation}
\ln \left( \frac{\Pi_{j}}{\Pi_{k}} \right) = 2\sigma_{\textrm{RM,interv}}^2 \lambda^4 \\.
\label{depoleqn}
\end{equation}

Following \citet{2011MNRAS.418.2336A}, the RM dispersion can be described in a simplified model of a turbulent magnetoionic medium as
\begin{equation}
\sigma_{\textrm{RM}}^2 \simeq \left( 0.81 \left< n_{e} \right> \left< B_{\textrm{turb}} \right> \right)^2 \frac{L d}{f} \\,
\label{turbeqn}
\end{equation}
where $n_{e}$ is the electron density in cm$^{-3}$ within the turbulent cells, $\left< n_{e} \right>$ is the average electron density in the volume along the path length traced by the telescope beam, $L$ is the path length in pc, $d$ is the size of the turbulent cells (also called the correlation length) in pc, $f$ is the filling factor of the cells given by $f=\left< n_{e} \right>/n_{e}$, and $\left< B_{\textrm{turb}} \right>$ in $\muup$G is the mean strength of the turbulent magnetic field, assumed to be the same both inside and outside of the cells. In particular, the characteristic scale of turbuence, $d$, the scale at which the dominant source of turbulence injects energy into the ISM of an LLS or DLA, is a key parameter to characterise magnetic turbulence in the associated gas. Finally it can be shown by combining equations \ref{depoleqn} and \ref{turbeqn} that
\begin{equation}
\frac{1}{2\lambda^4} \ln \left( \frac{\Pi_{j}}{\Pi_{k}} \right) \simeq  \left( 0.81 \left< n_{e} \right> \left< B_{\textrm{turb}} \right> \right)^2 \frac{L d}{f} \\,
\end{equation}
where all of the physical properties now parameterise the intervening absorption system, i.e.\ a DLA or LLS.

For the LLSs, we found a 95.5\% probability that there is a decrease in the polarized fraction relative to the control sample. From \citet{1999ApJ...511L..71P}, we use the LLS derived estimates for the electron density $=6.5\pm1.3\times10^{-2}$~cm$^{-3}$ and the path length $=3\pm1.6$~kpc, and assume a filling factor of 1. We previously obtained median polarized fractions of $4.2$\% for the controls and $3.3$\% for the LLSs, as observed at 1.4~GHz. We therefore derive $d \left< B_{\textrm{turb}} \right>^2 = 7.34$~pc~$\muup$G$^2$. This is equivalent to a typical turbulent scale of 7.34~pc in the LLSs for a turbulent magnetic field strength of 1~$\muup$G. However, we could realistically expect weaker turbulent fields: in Section~\ref{coherent} we estimated with 90\% probability that the regular coherent magnetic fields within the LLSs must be $\le2.4$~$\muup$G. If the random and coherent field components were of similar strength, this would increase the estimated turbulent scale to be on the order of $\approx17$~pc or larger. To summarise, to be consistent with the data, and realistic expectations for the magnetic field, the LLSs must have a turbulent scale on the order of $\approx5$--20~pc. This is similar to the characteristic outer scale of turbulence seen in normal galaxies, and is consistent with the case of e.g.\ galaxy NGC~6946, for which a turbulent scale of $20\pm10$~pc was estimated by \citet{1999ptep.proc....5B}, and is also in good agreement with the turbulent scale derived towards the Fan region in the Milky Way \citep{2013A&A...558A..72I}. The depolarization could also be conjectured to be the result of a patchy ionised medium in the LLSs, that serves as a depolarizing screen, in which case our estimates would constrain the physical size of the ionised clumps. However, the known high ionisation fraction of $\bar{x}\approx95$\% would appear to be inconsistent with this interpretation. We interpret our data as showing that an incoherent magnetic field must be present, and that the magnetised gas in LLSs must be highly turbulent.

For the DLAs, we found a 45.2\% probability that there is a decrease in the polarized fraction relative to the control sample. Depolarization by the DLAs therefore seems unlikely. However, if there is any decrease, we calculate a 90\% probability that a decrease relative to the control is of $\le0.80$\%. Using similar parameters to those derived in \citet{1999ApJ...511L..71P}, we calculate an upper limit for the DLAs, and a 90\% probability that $d \left< B_{\textrm{turb}} \right>^2 \le 6.52$~pc~$\muup$G$^2$. Depolarization can be caused by variations in the magnetic field, or in the thermal electron density. Due to self-shielding and the low ionised fraction, we do not expect any significant spatial variations in free electrons within the DLA environment. We interpret this to mean that there must be little variation in magnetic field within the DLAs. This must indicate either (i) that the disordered magnetic field strength is exceptionally weak, so that variations leave no measurable signal, or (ii) that the magnetic field must be highly coherent. Given the presumed young, dense, and clumpy form of the DLAs, together with our prior limits on the coherent magnetic field strength (see Section~\ref{coherent}), this favours the former interpretation. The weak random magnetic field strength suggests that the magnetised gas in DLAs is non-turbulent and quiescent.

\subsection{The Dynamo Paradigm}
Previous studies have been able to reliably observe the magnetised environment of strong Mg\,{\sc ii} absorbers \citep[e.g.][]{2014ApJ...795...63F}. We have now analysed the magnetic fields in two new types of absorption line systems, the DLAs and the LLSs, using modern polarimetric data. The magnetic properties of DLAs have only ever provided tentative results using 3--5 sources (see Section~\ref{introduction}), and in particular the magnetic environment of LLSs has never been studied before. It is therefore now of interest to consider the precise physical structures that are being probed at different stages of galaxy formation history, and to examine the constraints that are emerging for dynamo mechanisms \citep[e.g.][]{2015MNRAS.450.3472R}.

From \citet{2014ApJ...783L..20P}, it has previously been determined from cosmological simulations of the formation and evolution of galaxies that a prescribed tiny magnetic seed field grows exponentially through a small-scale dynamo until it saturates around $z = 4$ with a magnetic energy of about 10\% of the kinetic energy in the centre of the galaxy's main progenitor halo. By $z = 2$, a well-defined gaseous disk forms in which the magnetic field is further amplified by differential rotation, until it saturates at an average field strength of $\sim6$~$\muup$G in the disk plane. During this latter stage, the magnetic field is transformed from a chaotic small-scale field to an ordered large-scale field coherent on scales comparable to the disk radius.

Similarly \citet{2009A&A...494...21A} used dynamo theory to show that turbulence in protogalactic halos (a environment related to the LLSs, see Section~\ref{nature}) generated by thermal virialisation can drive an efficient turbulent dynamo, with the turbulent (small-scale) dynamo being able to amplify a weak seed magnetic field in halos of protogalaxies to a few $\muup$G strength within a few $10^8$ yr. Consistent with the cosmological simulations, this turbulent field served as a seed to the mean-field (large-scale) dynamo. Consequently, \citet{2009A&A...494...21A} found that galaxies similar to the Milky Way formed their disks at $z\approx10$, and regular fields of $\muup$G strength and a few kpc coherence length were generated within 2 Gyr (at $z\approx3$). However, field-ordering on the coherence scale of the galaxy size required an additional 6~Gyr (at $z\approx0.5$). Overall, this demonstrates that magnetic-field generation by the turbulent dynamo in galaxies requires neither large-scale rotation nor a disk, only turbulence. In addition, high-resolution simulations of protogalactic clouds have also demonstrated that significant turbulence can be generated prior to disk formation during the thermal virialisation of the halo \citep{2007ApJ...665..899W}. The big picture is then that strong turbulence in protogalactic halos at early epochs can drive the small-scale dynamo and amplify the seed field. This turbulent dynamo produces magnetic fields on scales comparable to the basic scale of galactic turbulence on timescales far shorter than that for the conventional mean-field galactic dynamo \citep[e.g.][]{1950RSPSA.201..405B}. In the epoch of disk formation, the turbulent field then served as a seed for the large-scale dynamo which develops in the disk of a newly formed galaxy.

Our data appear fully consistent with this conventional dynamo paradigm. We interpret our data as showing that the LLSs mostly consist of turbulent gas, while the DLAs mostly consist of quiescent gas. In the LLSs, we have found that turbulence has increased the random magnetic field, but not the coherent field. However, the presence of coherent fields may be just below our ability to detect them with these data (see Section~\ref{redshiftdist}). This is consistent with action from the small-scale dynamo, and with there having not been sufficient time for either disk formation or driving the large-scale dynamo. This has resulted in no, or a very limited, large scale field in LLSs. In contrast, in the DLAs, the gas appears to be mostly quiescent, and DLAs must have very weak non-detectable magnetic fields in both their coherent and random components. This is consistent with no detectable large-scale or small-scale magnetic field. This demonstrates limited, or a complete lack of, dynamo action in DLAs (see also Section~\ref{seedfields}). 

Our findings imply an evolutionary hierarchy, with different absorption lines probing varying stages of galaxy formation. The lack of a magnetic environment observed in DLAs, the random magnetic fields in LLSs, and the strong coherent fields observed in previous studies of strong Mg\,{\sc ii} absorbers \citep[e.g.][]{2014ApJ...795...63F}, all suggest that DLAs probe less-evolved galaxies than LLSs, and similarly that LLSs may be less-evolved counterparts to the strong Mg\,{\sc ii} absorbers. In this hierarchical scenario, the strong Mg\,{\sc ii} absorbers have had enough time to have driven the large-scale dynamo and to generate a coherent field. We are therefore exploring three different stages of magnetic field evolution. Notably, the only strong Mg\,{\sc ii} absorbers measured to date have been seen at lower redshifts of $z\approx1$, and are typically associated with normal star-forming galaxies \citep[e.g.][]{2014ApJ...795...63F} many of which have kinematics consistent with a disk/halo structure. In contrast, the to-date unstudied (in terms of magnetic field properties) strong Mg\,{\sc ii} systems at higher redshifts of $z\approx2$, are known to be associated with protogalactic structures \citep{2007ApJ...669..135M}. This would suggest that future studies of higher redshift strong Mg\,{\sc ii} absorbers may identify similar properties to those of the LLSs. This may hinder the use of QSO-absorption line systems for exploring the dynamo out to higher redshifts, in cases where the physical nature of the absorption line system evolves with time. The first step in this process will be further studies where we use Lyman systems out to $z\approx3$, and assume these will presumably evolve into normal galaxies by $z\approx0$, which is consistent with hierarchical galaxy formation evidenced by deep galaxy surveys. Multi-faceted studies that utilise many absorption lines, and explore the physical environment associated with each absorption line type, will be key to charting the evolution of magnetic fields throughout cosmic time. 

\subsection{Measuring the Magnetised Large-Scale Structure and Seed Fields}
\label{seedfields}
We have shown that the gas in DLAs is mostly quiescent, and that DLAs lack a detectable large-scale or small-scale magnetic field. This is consistent with no dynamo action having taken place in these protogalactic clumps. This implies that DLAs may therefore form natural repositories through which we can measure the magnetic field and infer the magnetic properties in the large-scale structure of the intergalactic medium (IGM) \citep[e.g.][]{2010ApJ...723..476A} and even possibly of magnetic seed fields. Observations with substantially larger samples and refined RM data via techniques such as $QU$-fitting \citep[e.g.][]{2012MNRAS.421.3300O} may be able to place tighter limits on these weak fields. However, direct detection of seed fields remains a very distant prospect. Observing a seed magnetic field of $10^{-6}$~$\muup$G requires the ability to detect a $\Delta\textrm{RRM}=0.66$~$\muup$rad~m$^{-2}$. The signal of interest is the difference in the standard deviation of the RM-distributions of the DLA, LLS, and control samples, where the intrinsic scatter due to QSO variations is $\sim15$~rad~m$^{-2}$. To first order, assuming we can calculate the standard deviation in RM to within an uncertainty of $\sigma/\sqrt{2(n-1)}$, we would require samples of $10^{15}$ polarized sources in order to measure such a signal. Such measurement is evidently beyond reach of even the SKA, for which $10^{7}$ RM measurements are expected during SKA phase 1 \citep{2015aska.confE..92J}. Such a measurement would also require substantial leaps in other areas: in the number of known DLAs, in Galactic RM foreground subtraction, and in the precision of RM measurements. 

However, DLAs may also represent intermediate stages of evolution, that lay somewhere between a dynamo-free clump and a small-scale dynamo fuelled LLS. This is highly likely as a DLA should have frozen in the magnetic fields from the IGM into its denser medium. In this case, the DLAs may still have turbulence and reasonable strength magnetic fields that are greater than a typical seed field, and possibly detectable with upcoming instruments. Such fields are currently below our detection threshold. Further data will be able to improve our estimates of the magnetic field in these objects.

\section{Conclusions \& Summary}
\label{conclusions}
We have studied samples of QSOs with intervening DLAs or LLSs identified somewhere along their line-of-sight, and compared these samples to a control sample of QSOs with no intervening DLAs or LLSs. We have compared these data using a Bayesian analysis, and it is the first time that such a statistical method has been applied to Faraday rotation and fractional polarization data in this way. The new application of this statistical tool opens a new method for quasar absorption line studies, in which many absorption lines have only a few quasar candidates and hence small sample sizes. This Bayesian analysis remains robust in the small sample regime, and therefore opens up previously restrictively-sized samples for analysis. This also presents a new way to more rigorously assess claimed correlations from previous studies of strong \citep[e.g.][]{2008Natur.454..302B,2014ApJ...795...63F} and weak \citep[e.g.][]{2016arXiv160400028K} Mg\,{\sc ii} absorbers.

Given our data, we have found that:
\begin{enumerate}
\item It is unlikely that DLAs have coherent magnetic fields, with a 32.1\% probability that DLA lines-of-sight have a higher RM than a control sample. It is also unlikely that DLAs have random magnetic fields, with a 45.2\% probability that DLA lines-of-sight have a lower polarized fraction than a control sample.
\item There is mild suggestive evidence that LLSs have coherent magnetic fields, and when controlling for the redshift distribution of our data, we find a 71.5\% probability that LLS lines-of-sight have a higher RM than a control sample. However, it is extremely likely that LLSs have random magnetic fields, with a 95.5\% probability that LLS lines-of-sight have a lower polarized fraction than a control sample.
\item We model our data to show that there is a 90\% probability that the regular coherent magnetic fields within the DLAs must be $\le2.8$~$\muup$G, and within the LLSs must be $\le2.4$~$\muup$G.
\item We also model the turbulent magnetic fields, and find that our data is most consistent with the DLAs having weak random magnetic fields which suggests that the magnetised gas in DLAs is non-turbulent and quiescent. We also find that the LLSs must have an incoherent magnetic field present, and that the magnetised gas must be highly turbulent with a turbulent scale on the order of $\approx5$--20~pc, which is similar to the measured turbulent scale within the Milky Way.
\end{enumerate}

Overall, a clear picture consistent with typical expectations of dynamo evolution is beginning to emerge from these absorption line studies. Our data are entirely consistent with the conventional interpretation of DLAs as protogalaxies, combined with the paradigm that dynamo action acts on weak disordered magnetic fields to generate coherent strong fields. Such studies hold promise for charting out dynamo evolution over cosmic time. Understanding the environment of different physical systems such as DLAs, LLSs, and other absorption lines will be first steps in beginning to directly map the evolution of cosmic magnetism. Using the combination of strong Mg\,{\sc ii} absorbers, LLSs, and DLAs, we have begun to chart out magnetic fields out to $z\approx2$. Our data allows us to see the first observational picture of magnetic field evolution in galaxies. Protogalaxies that lack a coherent magnetic field, and which maintain significant random magnetic fields, are evolving into normal star-forming galaxies with strong coherent fields. This reaffirms the role that magnetic fields play in the formation and evolution of galaxies. This is also the first time that the magnetic fields of LLSs have ever been studied, and the first time that DLAs or LLSs have been shown to have different magnetic field properties to those of a control sample. This also lends evidence for us to conclude that the seeding of magnetic fields by supernovae and subsequent amplification during structure formation are able to build up strong magnetic fields of $\muup$G strength within short timespans. This leads to detectable magnetic fields within the very first collapsing and starforming protohalos at high redshifts, which are the building blocks for the very first galaxies.

The developed techniques can also allow us to slowly push back towards the seed field and magnetogenesis era. Seed fields could have been generated in protogalaxies in the early Universe, e.g.\ at phase transitions, or in shocks in protogalactic halos (via the Biermann battery), or in fluctuations in the protogalactic plasma. Nevertheless, these seed fields will remain observationally out of reach for a long time, and remain 6 orders of magnitude below current observational thresholds, and so are expected to remain undetectable even with the SKA. However, with the introduction of a more refined Bayesian analysis for such studies, this work has created a quantitative gateway towards accessing and constraining magnetic fields in the IGM via quasar absorption line experiments. Future studies will be able to use the lowest column density Lyman systems, the Lyman-alpha forest at $\le10^{17}$~cm$^{-2}$, in order to constrain magnetism in the IGM.

We are currently lacking a large sample size in our study. This prevents us from subdividing the sample further, in order to explore other connections between the absorbers based on redshift, column density, and other intervenor parameters. In the future, with larger samples of sources with high polarized signal--to--noise so as not to be affected by Rician bias, we will be able to explore evolution based on the redshift of the absorbers, to definitively rule out evolution based on the redshift of the quasars, and to explore the evolution in magnetic and ionisation properties based on column density alone. Other subdivisions such as the spectral index are also of importance and has been shown to be important for studies of strong Mg\,{\sc ii} absorbers. Nevertheless, the spectral index is far less important for the DLAs and LLSs, which are known to be smoothly-distributed with a higher covering fraction compared to the clumpy partially-ionised medium with low-covering fraction that is associated with strong Mg\,{\sc ii}. It is therefore to be expected that the magnetic environment extends to larger scales in the DLAs and LLSs than in the Mg\,{\sc ii} absorbing systems. Furthermore, the DLA and LLS sample also include a very tight cross-matching criterion of 2~arcsec, which is equal to 25~kpc at $z=1$, and therefore less than typical estimates for the DLA/LLS size. Increased sample sizes with higher angular resolution will allow us further subdivide our data, and to test these attributes. Future studies will therefore be able to subdivide the data into flat- and steep- samples with high angular resolution, allowing for direct tests of the requirement for single-component compact sources, and also providing a unique way to measure the ionised size of these absorbers. Future data that will become available as DLAs and other absorption features are identified in the SDSS DR10 or DR12 will also improve our sample size, and allow for further subdivision into smaller sub-populations \citep{2016arXiv160806483P,2016arXiv160805112R}.

Finally, future studies will also be able to refine the Bayesian analysis of Faraday rotation data introduced here for quasar absorption lines. Testing other distributions, alongside the normal distributions assumed in this paper, will provide improvements and will refine future results with larger samples. Such an analysis will most likely be unable to use conjugate priors, and so will require more sophisticated sampling techniques to be used. We also discuss our model selection in Appendix~\ref{modelchoice}. If there is substantial deviation from these model choices, then there would be implications for most previous magnetic field studies throughout the literature, as our chosen models have often been used either explicitly via model-fitting or the calculation of standard deviations, or implicitly via the use of frequentist tests. While we have been able to model the coherent magnetic fields using the RMs and GRMs, further analysis of the distribution of these statistical samples will allow us to ensure that our implicit assumption, that $\sigma_{\textrm{RM}}^2 = \sigma_{\textrm{GRM}}^2 + \sigma_{\textrm{RRM}}^2$ (see Section~\ref{faradaybayes}), is correct. Expansion of our Bayesian technique may also allow us to incorporate modelling of the error terms that currently inhibit use of RRMs, and thereby allow a statistical measurement with the RRMs directly. While this could have implications for our coherent magnetic field estimates in both DLAs and LLSs, our primary result of depolarization associated with the LLSs would of course remain unaffected. The application of these Bayesian techniques offers the opportunity to remove the systematic effects that frequently affect correlation-based studies, and to provide a thorough rigourous framework for testing connections between other absorption line systems.







%
%
%
\section*{Acknowledgements}
The National Radio Astronomy Observatory is a facility of the National Science Foundation operated under cooperative agreement by Associated Universities, Inc. This research has made use of the SDSS, for which funding has been provided by the Alfred P. Sloan Foundation, the Participating Institutions, the National Science Foundation, and the U.S. Department of Energy Office of Science. B.M.G. acknowledges the support of the  Natural Sciences and Engineering Research Council of Canada (NSERC) through grant RGPIN-2015-05948. The Dunlap Institute is funded through an endowment established by the David Dunlap family and the University of Toronto. SPO acknowledges support from UNAM through the PAPIIT project IA103416.


\newpage
\appendix
\section{Bayesian Analysis}\label{appendix-bayes}
%
%
\subsection{Analysis of Faraday Rotation}
\label{faraday-appendix}
We model the distribution of RMs and GRMs with Gaussian distributions centred at 0~rad~m$^{-2}$. Using the RMs and GRMs, we therefore wish to model the joint posterior distribution of both the mean, $\mu$, and variance, $\sigma^2$, given the data, $D$, for which the posterior is given by $p(\mu,\sigma^2|D)$.

We apply Bayes' theorem, using the notation of \citet{gelman2004} throughout,
\begin{equation}
p(\mu,\sigma^2|D) = \frac{p(D|\mu,\sigma^2)p(\mu,\sigma^2)}{p(D)} ,
\end{equation}
where $p(D|\mu,\sigma^2)$ is the Gaussian likelihood function, $p(D)$ is purely a function of the data, and $p(\mu,\sigma^2)$ is the joint prior distribution. The probability of the data, $p(D)$, is just a normalisation constant and does not affect our results. 

\begin{figure}
\centering
\includegraphics[clip=false, width=\hsize]{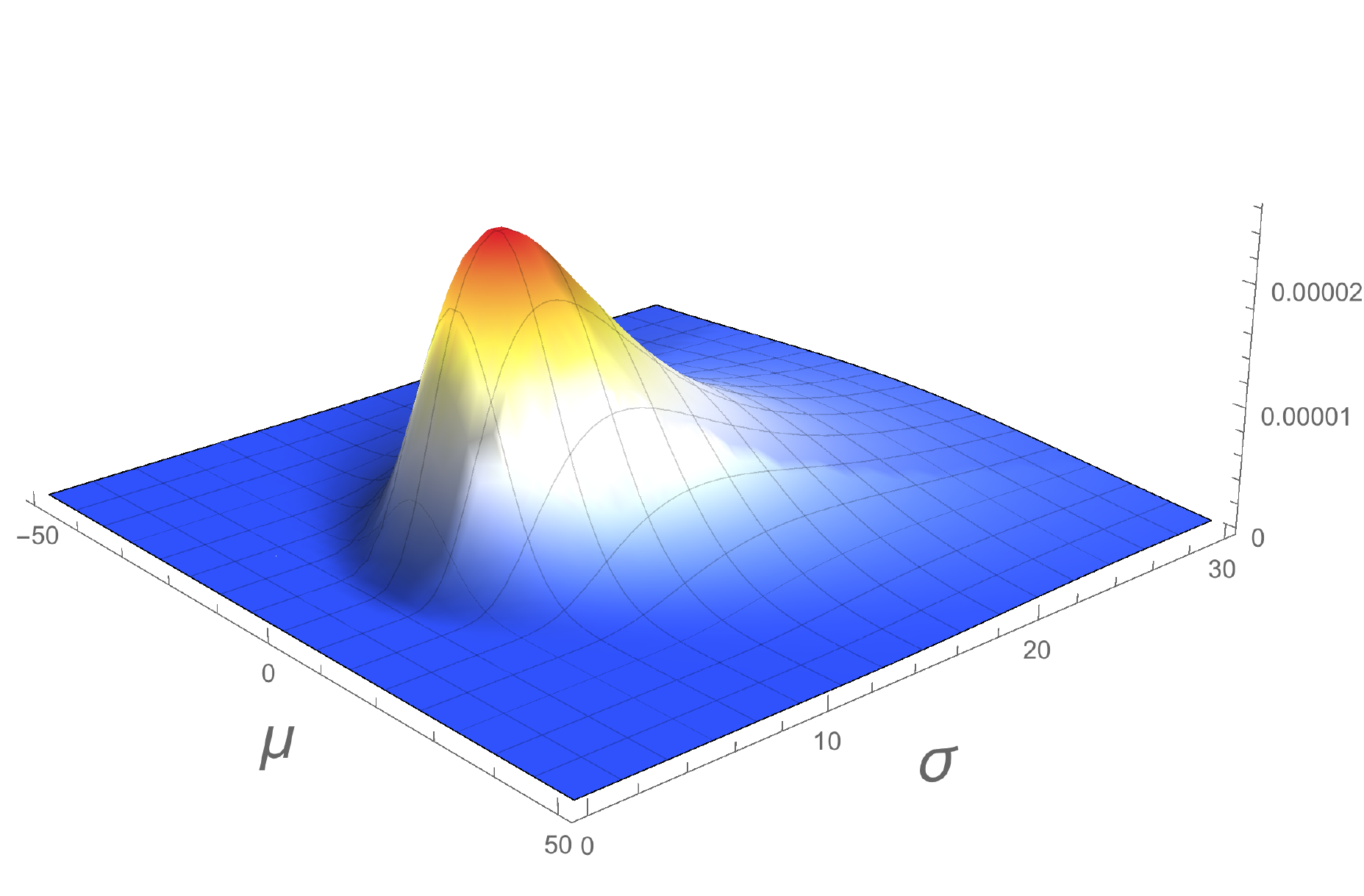}
\includegraphics[clip=false, width=\hsize]{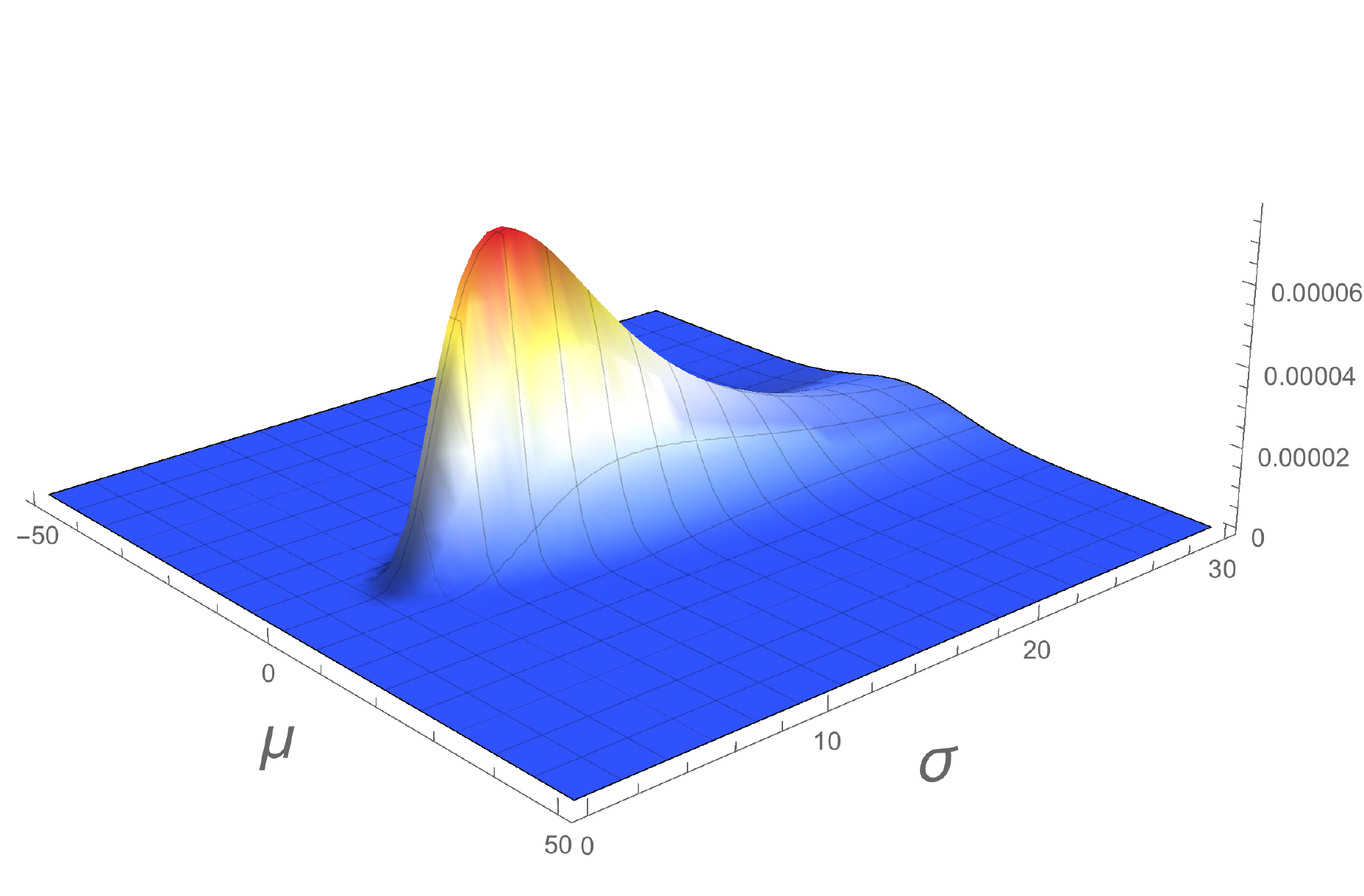}
\caption{\small{The selected priors used for the Faraday rotation estimates. Both the RM (top) and GRM (bottom) priors are shown. $\mu$ and $\sigma$ both have units of rad~m$^{-2}$. Both priors are weakly-informative, and we confirm that we are not sensitive to the choice of prior in Section~\ref{robustness}.}}
\label{priorRM}
\end{figure}

Following \citet{gelman2004}, we choose a conjugate prior distribution for the two-parameter univariate normal sampling model, as this has the advantage that it is analytically tractable. The conjugate prior is of the form $p(\mu,\sigma^2)=p(\sigma^2)p(\mu|\sigma^2)$, where $p(\sigma^2)$ is a scaled inverse-$\chi^2$ distribution and $p(\mu|\sigma^2)$ is a normal distribution. This corresponds to the joint prior distribution,
\begin{equation}
p(\mu,\sigma^2) \propto \sigma^{-1} (\sigma^2)^{-(1+\nu_{0}/2)} \exp{\left(-\frac{1}{2\sigma^2} \left[\nu_{0}\sigma_{0}^2 + \kappa_{0}(\mu_{0}-\mu)^2 \right] \right)} ,
\end{equation}
where the four hyperparameters can be identified, in simple terms, as $\mu_{0}$ -- the prior mean, $\kappa_{0}$ -- our degree of belief in this parameter (which is also equivalent to the number of prior measurements, while the scale of $\mu_{0}$ is given by $\sigma_{0}^2/\kappa_{0}$), $\sigma_{0}^2$ -- the prior variance, and $\nu_{0}$ -- our degree of belief in this parameter (equivalent to the degrees of freedom of $\sigma_{0}^2$). We choose our prior parameters for the RM to be weakly-informative, setting $\mu_{0} = 0.0$~rad~m$^{-2}$, $\kappa_{0} = 1.0$, $\sigma_{0}^2 = 25.0^2$~rad$^2$~m$^{-4}$, and $\nu_{0} = 0.5$. For the GRM, we also use a weakly-informative prior but include mildly stronger belief that the mean of the GRM is centred at 0.0~rad~m$^{-2}$, setting $\mu_{0} = 0.0$~rad~m$^{-2}$, $\kappa_{0} = 10.0$, $\sigma_{0}^2 = 25.0^2$~rad$^2$~m$^{-4}$, and $\nu_{0} = 0.5$. The priors for the Faraday rotation estimates are shown in Fig.~\ref{priorRM}.

\subsection{Analysis of Polarized Fractions}
\label{polfrac-appendix}
We now consider the case of the polarized fraction values, $\Pi$, associated with the DLA and LLS. Unlike the RM and GRM values, the $\Pi$ values are likely not well-modelled by a normal distribution. As we know that the polarized fractions are positive-definite, and also that the distributions appear skewed towards zero (see Fig.~\ref{pihistograms}), we instead select a log-normal distribution as our model \citep[e.g.][]{2012AdAst2012E..52T,2013MNRAS.436.2915M}.\footnote{A log-normal distribution is a distribution whose logarithm is normally distributed.}
%

We therefore first take the logarithm of the data samples (as shown in Fig.~\ref{pihistograms}), and our approach is then similar to that used in Section~\ref{faradaybayes} for normal data. Again using a conjugate prior and a Gaussian likelihood, we specify posterior distributions of the parameters $\mu$ and $\sigma$. We then draw samples of $\mu$ and $\sigma$ from the posteriors, and transform each of them back into non-logarithmic space. As we are interested in the arithmetic mean and arithmetic standard deviation of each of the log-normal distributions, these are given by,
\begin{equation}
\textrm{E}\left[\Pi\right] = e^{\mu+\frac{\sigma^2}{2}} ,
\end{equation}
and,
\begin{equation}
\textrm{SD}\left[\Pi\right] = e^{\mu+\frac{\sigma^2}{2}}\sqrt{e^{\sigma^2}-1} .
\end{equation}

\begin{figure}
\centering
\includegraphics[clip=false, width=\hsize]{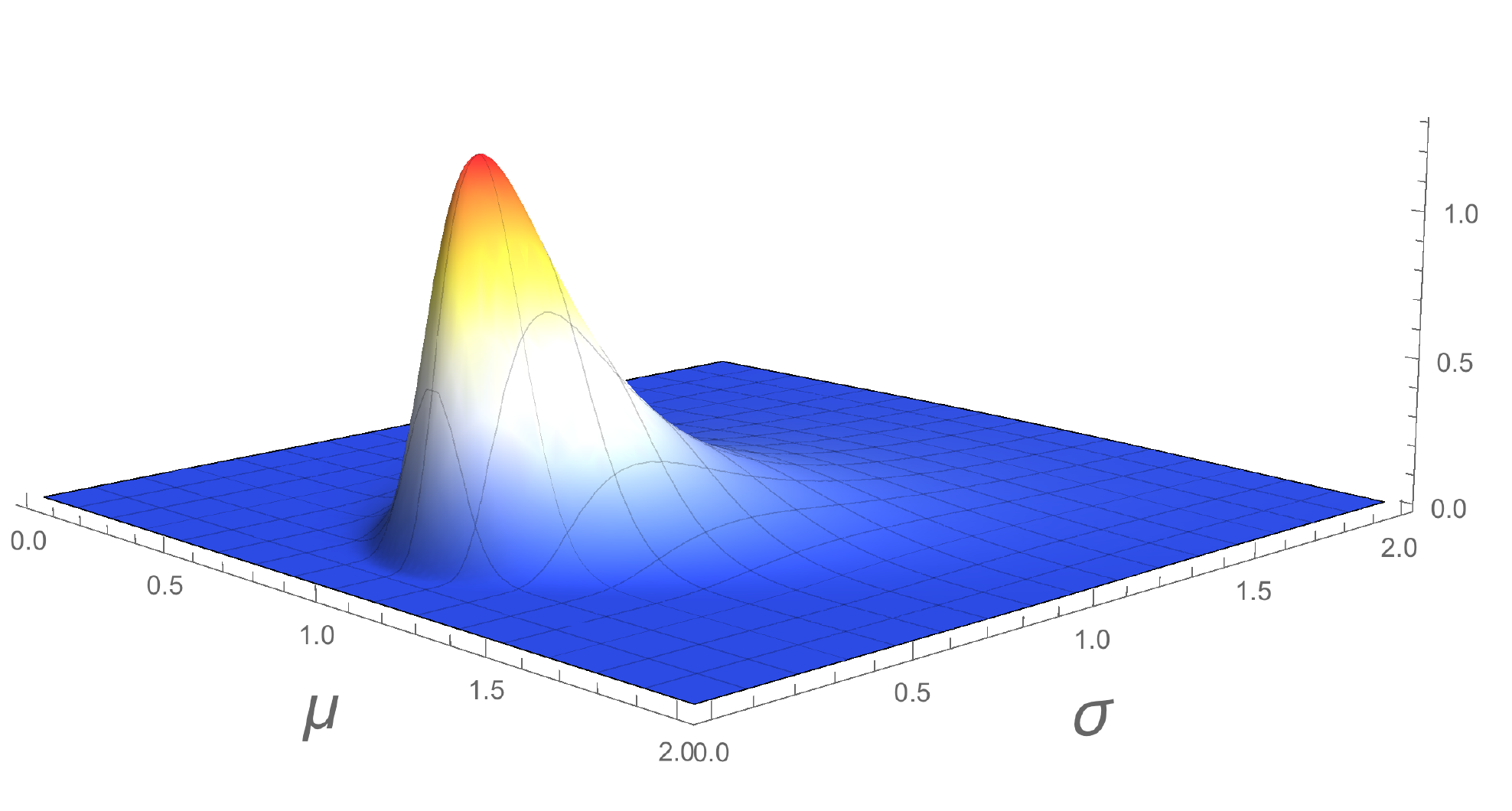}
\caption{\small{The selected prior used for the polarized fraction estimates. $\mu$ and $\sigma$ are both unitless. The prior is weakly-informative, and we confirm that we are not sensitive to the choice of prior in Section~\ref{robustness}.}}
\label{priorPI}
\end{figure}

The full-framework is otherwise done in the same manner as in Section~\ref{faraday-appendix}. We choose our prior parameters to be weakly-informative, setting $\mu_{0} = 1.0$, $\kappa_{0} = 10.0$, $\sigma_{0}^2 = 1.0^2$, and $\nu_{0} = 0.5$. $\mu$ and $\sigma$ are both unitless. The prior is shown in Fig.~\ref{priorPI}.

\subsection{Drawing Samples from the Posterior}
\label{drawing-appendix}
Also following \citet{gelman2004}, as we have chosen a conjugate prior, we have a closed-form expression for the joint posterior distribution, which is given by,
\begin{equation}
p(\mu, \sigma^2|D) = \textrm{N-Inv-}\chi^2 \left( \mu_{n}, \sigma^2_{n}/\kappa_{n}; \nu_{n}, \sigma^2_{n} \right).
\end{equation}
where $\textrm{N-Inv-}\chi^2$ is the normal-inverse-$\chi^2$ distribution.

The conditional posterior distribution, $p(\mu|\sigma^2,D)$, of $\mu$, given $\sigma$, is then proportional to the joint posterior distribution with $\sigma^2$ held constant, which is given by,
\begin{equation}
\mu|\sigma^2,D \sim \textrm{N}\left( \mu_{n}, \sigma^2/\kappa_{n} \right) ,
\label{conditional}
\end{equation}
where $\textrm{N}$ is the normal distribution, and while the marginal posterior distribution, $p(\sigma^2|D)$, of $\sigma^2$, is a scaled inverse-$\chi^2$, which is given by,
\begin{equation}
\sigma^2|D \sim \textrm{Inv-}\chi^2\left( \nu_{n}, \sigma^2_{n} \right) .
\label{marginal}
\end{equation}
In order to sample from the joint posterior distribution, we will first draw a $\sigma^2$ from the marginal posterior distribution of equation~\ref{marginal}, and then draw $\mu$ from the normal conditional posterior distribution of equation~\ref{conditional}, using the simulated value of $\sigma^2$. In practice, we will instead draw samples using the inverse-$\Gamma$ (Inv-$\Gamma$) distribution, and using the identity for an Inv-$\chi^2\left(\nu_{0}, \sigma_{0}^2 \right)$ $\equiv$ Inv-$\Gamma\left( \nu_{0}/2, \nu_{0}\sigma_{0}^2/2 \right)$. We will sample 2,000,000 times in order to generate our samples of $\mu$ and $\sigma$.

\subsection{Model Selection}
\label{modelchoice}
To quote George Box, ``all models are wrong, but some are useful''. Our model selection is justified as our data show direct similarity to these distributions, and as the models we use have been widely employed in previous studies with larger sample sizes \citep{2012arXiv1209.1438H,2012AdAst2012E..52T,2013MNRAS.436.2915M,2014ApJS..212...15F}. If our assumed normal- or log-normal distribution is a very bad approximation, then this would affect the conclusions of our paper. More importantly, if our models are incorrect, then this would have wide implications for the conclusions of various papers across the literature. 

In the future, with larger samples, we will be able to more directly test our models, to expand upon our Bayesian framework to include an error analysis (possibly allowing for use of RRMs), and to test the effects of assuming different model distributions. These more developed cases will likely not be able to use a conjugate prior, and will require more sophisticated techniques for sampling from the posterior distribution.





\end{document}